\documentclass[journal=jcisd8,manuscript=article]{achemso}

\usepackage[T1]{fontenc} 
\usepackage{url}
\usepackage{longtable}
\usepackage{graphicx}
\newcommand{\etal}{\textit{et al}.}

\author{Maho Nakata}
\affiliation[RIKEN ACCC]
{Head Office for Information Systems and Cybersecurity, RIKEN, 2-1, Hirosawa, Wako-City, Saitama, 351-0198 JAPAN}
\email{maho@riken.jp}
\author{Tomomi Shimazaki}
\affiliation[KOBE]
{Kobe University, Graduate School of System Informatics, Rokkodai-cho 1-1, Nada-ku, Kobe, Hyogo 657-8501, Japan.}
\author{Masatomo Hashimoto}
\author{Toshiyuki Maeda}
\affiliation[STAIR]
{Software Technology and Artificial Intelligence Research Laboratory,\\Chiba Institute of Technology, 2-17-1 Tsudanuma, Narashino, Chiba 275-0016, JAPAN}

\title{PubChemQC PM6: A dataset of 221 million molecules with optimized molecular geometries and electronic properties }

\begin{document}

\begin{abstract}
We report on the largest dataset of optimized molecular geometries and electronic properties calculated by the PM6 method for 92.9\% of the 91.2 million molecules cataloged in PubChem Compounds retrieved on Aug. 29, 2016. In addition to neutral states, we also calculated those for cationic, anionic, and spin flipped electronic states of 56.2\%, 49.7\%, and 41.3\% of the molecules, respectively. Thus, the grand total calculated is 221 million molecules. The dataset is available at \url{http://pubchemqc.riken.jp/pm6_dataset.html} under the Creative Commons Attribution 4.0 International license.
\end{abstract}

\section{1. Introduction}
The importance of exploring new organic molecules is increasing for the development and design of organic thin film solar cells~\cite{brabec_hauch_schilinsky_waldauf_2005}, electroluminescent materials~\cite{Kaji2015}, organic nonlinear optical materials~\cite{B512646K}, molecular sensors~\cite{VALEUR20003}, and new drugs~\cite{Drugdesign}. Ab initio quantum chemical calculations are useful for exploring such organic molecules as they give accurate predictions for the chemical properties without time-consuming physical\slash chemical experiments~\cite{Helgaker2000}. However, there are also problems. Even though the computational cost of the quantum chemical calculations has been reduced considerably, they are still too slow to explore the chemical space of the molecular compounds. For example, the number of drug candidates is estimated to be $10^{60}$, assuming the Lipinski rule~\cite{LIPINSKI19973,chemicalspace}. Consequently, we must depend on empirical methods.

Machine learning is a promising empirical method for chemistry, and it is frequently applied to molecules and solids~\cite{1903.04388,Butler18:_machin,PhysRevLett.108.253002,PhysRevLett.108.058301,Butler2018,Hansen2015,Bartoke1701816,Ramprasad2017,ml1,ml2,ml3,Gomez-Bombarelli18:_automa}. To apply a machine learning technique, we require a lot of high-quality training data. Unfortunately, not so many datasets for virtual screening are readily available on the Internet~\cite{Nist2005,GDB-17,QC134K,1703.00564,Smith17,Ghahremanpour18,Pereira17,PubChemQC2015,PubChemQC2017}. As such, there is a need to perform quantum chemical calculations to provide training data on molecular geometries, electronic structures, and other properties.

To provide this training data, we have to consider two major issues: the first is a criterion for choosing molecules; the second is a representation of molecules.

Regarding the criterion for choosing molecules, there are an astronomical number of molecules, even if we restrict the atomic species and number of atoms. For example, Ruddigkeit \etal{} created a dataset called GDB-17~\cite{GDB-17}, which enumerates 166 billion organic small molecules consisting of up to 17 atoms of C, N, O, S, and halogens.

The problem is that it is not straightforward to judge which molecules are essential. For example, Ramakrishnan \etal{} provide the QM7, QM7b, QM8, and QM9~\cite{QC134K,GDB-17} datasets, which are subsets of GDB-17 wherein the geometries and electronic structures are calculated using quantum chemical methods. The QM9 dataset is the largest among them with 134 thousand molecules (these datasets are included in the Moleculenet~\cite{1703.00564} as well). Unfortunately, such an enumeration algorithm may not work even when the sizes of the molecules become only slightly more abundant ; there are 27,711,253,769 isomers for $\rm C_{32}$~\cite{wikipedia_alkane}, and almost all of them might be insignificant. Another example would be cis-trans isomerism. There are $2^N$ isomers for $N$ cis-trans double bonds. Choosing all the isomers is unnecessary; only representable isomers should be selected. Moreover, sometimes these isomers function very differently, while other times they work very similarly.  For instance, cis-unsaturated fatty acids can promote good cholesterol, whereas trans fatty acids are considered harmful~\cite{transfat}. On the other hand, 1,3-dichloropropene is used as an agricultural chemical, and there is no significant difference in cis- and trans-1,3-dichloropropene. Usually, a mixture is used~\cite{dichloropropene}. Thus, it is not straightforward to decide which isomers to choose or whether to choose both. 

Therefore, instead of choosing molecules by ourselves, we decided to develop a chemical database that is believed to have essential molecules and perform quantum chemical calculations on the molecules in the database as much as possible. There are many chemical databases on the Internet, including CAS~\cite{CAS}, ChEMBL~\cite{chembl}, ChemSpider~\cite{chemspider}, Zinc~\cite{zinc}, and PubChem~\cite{10.1093/nar/gky1033}. Here, we chose PubChem Compounds~\cite{10.1093/nar/gky1033} as the reference set of compounds as it is one of the most extensive and comprehensive chemical compound databases, containing approximately 97 million molecules. Its records are from hundreds of large data sources including those of universities, pharmaceutical companies, government agencies, chemical vendors, scientific papers, and other curation efforts. We believe PubChem Compounds includes most of the essential molecules as well as a sufficient variety of them.

As for the second issue of a representation of molecules, it is worth noting that, interestingly, there is no rigorous definition of a molecule. Nevertheless, we can define a molecule under certain assumptions. In the gas phase under Born-Oppenheimer approximation, non-relativistic limit, and point charge nucleus model, we can determine the Hamiltonian of a molecule by a set of atoms with Cartesian coordinates, and the number of electrons in the system. Then, we can solve the Schr\"odinger equation to obtain the wavefunctions and the quantum numbers. In this way, a molecule is defined by the Hamiltonian, the wavefunction, and the quantum numbers. A problem with this definition is that we cannot easily distinguish two different molecules. On the other hand, although the most convenient representation would seem to be a common name, someone would have to name each new compound. Here, we can employ IUPAC nomenclature~\cite{IUPAC} as a systematic nomenclature of molecules, since atoms, connectivity of atoms, bond orders, and other stereo information are enough to specify a molecule in most cases. Nevertheless, we cannot easily process IUPAC names on computers. 

Therefore, we decided to rely on human and machine-readable molecular encoding systems such as InChI (International Chemical Identifier)~\cite{Heller2015,INCHI1,INCHI2} and SMILES (Simplified Molecular-Input Line-Entry System)~\cite{smiles1,smiles2}. These are mostly compatible  with IUPAC nomenclature, and we can encode a molecule like a chemical formula in a systematic way. Both InChI and SMILES encode the compounds in PubChem Compounds. We made extensive use of them in the calculations; in particular, we used SMILES for generating the initial geometry guess and InChI for validating the optimized results. Note that these encodings have some ambiguities; they only define a set of atoms with electronic charge and some information about the three-dimensional configuration of the atoms.

According to the above considerations, we have been developing datasets by performing quantum chemical calculations~\cite{PubChemQC2015,PubChemQC2017}. In this paper, we report on our development of a dataset by performing geometry optimization and calculating the electronic structure and other properties by using the PM6 method~\cite{Stewart2007} on molecules listed in PubChem Compounds on the basis of SMILES and InChI encodings. We call the dataset the \emph{PubChemQC PM6 dataset}. We used the PubChem Compounds dataset retrieved on Aug. 29, 2016. It consists of 91,679,247 molecules. First, we excluded molecules whose molecular weights are greater than 1000g/mol (0.66\%) and charged (ionized) molecules (2.39\%). Then, we performed geometry optimization on each molecule. In addition to the neutral state, we considered cationic, anionic, and spin flipped states for each compound as well. At the time of writing, we have successfully calculated for 86,213,135 neutral states, 51,555,911 cationic states, 45,581,750 anionic states, 37,839,619 spin flipped states. Thus, the grand total calculated is 221,190,415 molecules of optimized geometries and electronic structures. 

The coverage of PubChemQC PM6 is over 94.0\% for neutral molecules. To the best of the authors' knowledge, this is the largest dataset developed by semi-empirical quantum chemical calculations. More specifically, the number of records in PubChemQC PM6 is greater than that in any other dataset that employed any quantum chemical calculation, ignoring the differences in calculation methods and their calculation accuracy. Examples of the datasets created using the density functional theory (DFT) method, which is more accurate than the PM6 semi-empirical method, are the Harvard Clean Energy Project Database~\cite{C3EE42756K,Hachmann11} and the ANI-1 dataset~\cite{Smith17}. The former has 2.3 million candidate compounds for organic photovoltaics and the latter contains 20 million calculated off-equilibrium conformations for organic molecules; that is, they are smaller than PubChemQC PM6.

The dataset is compressed and available at \url{http://pubchemqc.riken.jp/pm6_dataset.html} under a Creative Commons Attribution 4.0 International License.

The rest of the paper is organized as follows. Section 2 briefly introduces the PubChem project and two molecular encodings: InChI and SMILES. Section 3 discusses the validity of the PM6 calculations. The workflow for developing our database, calculation results, availability, and explanation of the PubChemQC PM6 dataset are shown in Section 4. Section 5 discusses our results, and Section 6 describes future work.

\section{2. PubChem Compound dataset and molecular encodings}
\label{sec:pubchem}
PubChem is an open chemistry database maintained by the National Institutes of Health (NIH). Since 2004, it has been continually updated with molecular information from all over the world. It consists of three sub-databases: Compounds, Substances, and BioAssays. While Substances just archives the submitted data, Compounds contains unique standardized compound data extracted from Substances. PubChem Compounds contains 97,125,741 compounds as of Feb. 14, 2019. Each compound record in Compounds has an InChI and SMILES as well as a PubChem CID (unique compound ID assigned by PubChem), molecular formula, and other information. In the rest of the paper, we will refer to the PubChem CID as CID.

Both InChI and SMILES give a human and machine readable representation in the form of an ASCII string for each molecule. They each have their pros and cons, so we decided to take the best of both.

An InChI representation is a sequence of layers prefixed by its version number. The layers are main, charge, stereochemical, isotopic, fixed-H, and reconnected. The main layer consists of a chemical formula, atom connections, and hydrogen atoms. The charge layer contains information about protons and charges. Since multiple representations are possible for a molecule, InChI defines an injective standardization procedure so that two different molecules will have different standardized InChIs. Such standardization is vital to verify the calculation results. For example, the standard InChI representation of ethanol (CID 702), is the following:
\begin{verbatim}
InChI=1S/C2H6O/c1-2-3/h3H,2H2,1H3
\end{verbatim}
\texttt{InChI=1S} stands for standardized InChI. \texttt{C2H6O} is the chemical formula of ethanol. \texttt{c1-2-3} shows how carbons are connected. How many hydrogens are connected to each of the other atoms are represented by \texttt{h3H,2H2,1H3}. For another example, L-ascorbic acid (CID 54670067) is described as follows.

\begin{verbatim}
InChI=1S/C6H8O6/c7-1-2(8)5-3(9)4(10)6(11)12-5/h2,5,7-8,10-11H,1H2/t2-,5+/m0/s1
\end{verbatim}
Full documents and InChI Software source codes can be found on the original website~\cite{INCHI2,Heller2015}.

SMILES notations have better human readability and are more popular than those of InChI. Unfortunately, however, the original SMILES is a proprietary format, and its details are closed. Moreover, there exist many dialects derived from the original SMILES, while several standardization methods are available. Among those, we employ a canonicalization of SMILES based on InChI proposed by O'Boyle~\cite{boyle2012} and implemented it in Open~Babel \cite{OpenBabel}. In SMILES, ethanol is represented as follows.
\begin{verbatim}
CCO
\end{verbatim}
Note that \texttt{C(O)C} also represents ethanol, but it is not canonical as the description of the atomic connectivity is not unique. L-ascorbic acid has a SMILES representation:
\begin{verbatim}
C(C(C1C(=C(C(=O)O1)O)O)O)O
\end{verbatim}
It has an isomeric description used in PubChem Compounds:
\begin{verbatim}
C([C@@H]([C@@H]1C(=C(C(=O)O1)O)O)O)O
\end{verbatim}
By canonicalizing with Open~Babel, we have the following:
\begin{verbatim}
OC[C@@H]([C@H]1OC(=O)C(=C1O)O)O
\end{verbatim}

PubChem Compounds provides canonical SMILES and isomeric SMILES. Canonical SMILES does not include information about stereoisomerism, while isomeric SMILES provides relevant geometric details. Open~Babel can handle canonical SMILES and OpenSMILES. The former canonicalizes the SMILES representation via InChI standardization.

The pros and cons of the InChI format (thus, the cons and pros of the SMILES format) are summarized as follows.

\noindent\textbf{Pros}
\begin{itemize}
\item It is standardized by IUPAC. No variants are allowed, unlike SMILES.
\item The standardization algorithm is unique and publically available.
\item The treatment of hydrogens, charges, and isotopes is much more systematic than in SMILES.
\end{itemize}
\noindent\textbf{Cons}
\begin{itemize}
\item It is less human readable than SMILES.
\end{itemize}

Every molecular encoding has its limitations. The above encodings each have ambiguities. We define a molecule in the gas phase under Born-Oppenheimer approximation, non-relativistic limit, and point charge nucleus model as follows. The total Hamiltonian of the system is determined by the Cartesian coordinates of atoms and the number of electrons. Then, a molecule is defined as the wavefunction and its quantum numbers in the solution of the Schr\"odinger equation. Clearly, InChI and SMILES have ambiguities; they do not explicitly include Cartesian coordinates, and instead have only a three-dimensional configuration of molecules. The bond order and formal charge in SMILES are empirical parameters that do not appear in the Hamiltonian. Moreover, no rigorous conversion between them is available; those of SMILES are proprietary, and there are many dialects. This has led many different SMILES for a compound~\cite{InChIFAQ}, and bond order is not supported in InChI. 

On the other hand, converting common names to a SMILES or InChI representation requires a large table. There is a trade-off between specificity and generality, and we believe that InChI and SMILES are the best choices for the present.

\section{3. Validity of PM6 calculation for molecular geometry optimization}
\label{sec:pm6}
PM6 is a semi-empirical method that neglects the diatomic differential overlap approximation developed by Stewart~\cite{Stewart2007}.  PM6 is a promising semi-empirical method for geometry optimization. It supports broad types of 83 elements including H-Ba and Lu-Bi, except for lanthanoids and actinoids. In addition, it is capable of reproducing proper bond lengths and angles. Suppose that we use H, C, N, O, F, P, S, Cl, Br, and I. Then, the average unsigned errors in bond length is 0.031 angstroms, and the average unsigned errors of bond angles is 3.2 degrees. Although PM6 has a younger sibling, PM7~\cite{Stewart2013}, PM6 is known to give slightly better results~\cite{PM6PM7comp,PM6accuracy} in terms of heat of formation, bond lengths, dipole moments, and ionization potential. For a set of approximately 5000 molecules, Stewart showed the average unsigned error of bond length calculations made by PM6 was 0.087 angstrom, while that of PM7 was 0.098. Moreover, for another set of similar size, he showed that the average unsigned errors of dipole moments by PM6 was 0.82 debye, and that by PM7 was 1.08 debye.

Stewart further showed that the average unsigned error of bond length and bond angle calculations for 70 elements by PM6 were 0.091 angstroms and 7.9 degrees for 70 elements. He concluded that the overall accuracy of PM6 in predicting heats of formation for compounds of interest in biochemistry is somewhat better than B3LYP/6-31G*. Note that B3LYP geometry optimization processes usually provide quite good geometries; typical errors of bond angles and bond length are known to be within a few degrees and 0.02 angstrom ~\cite{handbookcc,performanceofb3lyp631gd,performanceofb3lyp631gd_2,performanceofb3lyp631gd_3}.

Since B3LYP is more faithful to the law of physics, the quality of the results of the B3LYP method is better than those of PM6, and thus it might be preferable for our purpose. However, we observed that it took a year for B3LYP to perform geometry optimizations of only one million molecules, while it took almost the same amount of time for PM6 to accomplish that for 100 times more molecules. In terms of processing speed of geometry optimization, we prefer PM6 as it achieves a better trade-off between speed and quality than B3LYP. Nonetheless, we can perform B3LYP calculations by making use of the geometry optimization results of PM6. Since PM6 optimized geometries are fairly good, the B3LYP one point calculation would also give good electronic structures.

\section{4. PubChemQC PM6 dataset}
\label{sec:pubchemqc}
We downloaded all of the molecular SDF (structure-data file) from the PubChem ftp site on Aug. 29, 2016. Then, we parsed them and extracted molecular record containing CID (Compound ID), molecular weight, InChI representation, isomeric SMILES, molecular composition formula, electronic charge, and spin number for each molecule. We calculated the electronic charge from the isomeric SMILES representation of the molecule. We set the spin number to 0 or 1 according to the parity of the number of electrons in the system. Then, we sorted the records by molecular weight in ascending order and excluded molecules with molecular weights larger than 1000g/mol. This molecular weight limit is larger than the Lipinski rule 500g/mol~\cite{LIPINSKI19973}, and there were only 604,330 such molecules (0.66\%) in PubChem Compound.  We also excluded charged 2,188,881 (2.39\%) molecules. The input files were generated by Open~Babel\cite{OpenBabel} using the isomeric SMILES representation of the molecule with the \texttt{--addh} and \texttt{--gen3d} options. In our experience, Open~Babel generates somewhat more reasonable initial geometry guesses for SMILES than for InChI. One reason might be that the bond angle estimation is easier for SMILES as it records the bond order, while InChI does not record it. Next, we calculated the PM6 optimized geometry of the molecules by using Gaussian09~\cite{gaussian09}. If it succeeded, we also performed the geometry optimization for the cationic state, anionic state, and spin flipped state using the PM6 optimized geometry as the initial geometry guess. Finally, we checked that the calculated InChI coincided with the original InChI at the optimized geometry by using Open~Babel of the neutral state.  We verified the identity of the chemical formula and atom connection of the main layer. In particular, we used the following \texttt{sed} script for the two InChI representations:
\begin{verbatim}
sed -i -e 's|/[abd-z].*$||g'
\end{verbatim}
and verified the identity of the outputs. We ignored the other layers that include floating hydrogens, formal charge on atoms, and total charge. Because we cannot calculate or find meanings of them from the outputs of quantum chemical calculations. We also ignored stereoisomers, geometric isomers, and conformers.

The resultant data for each CID are as follows: input file, atom coordinates in xyz format, and a JSON file. We parsed the output files from Gaussian09~\cite{gaussian09} with cclib~\cite{cclib} to generate the JSON files. We trimmed ``displace'' sections from the JSON files. We summarized the calculation status (successful or not) in a MySQL database. See the following section for the success rate. We heavily used the GNU parallel utility~\cite{tange_ole_2018_1146014} to perform almost all of the calculations.

All of the calculations were performed on The RIKEN HOKUSAI BigWave supercomputer (Intel Xeon Gold 6148 2.4GHz, 1,680 CPUs, 33,600 cores), QUEST cluster (Intel Core2 L7400 1.50 GHz, 700 nodes, 1400 cores), and RIKEN RICC supercomputer (Intel Xeon 5570 2.93 GHz, 1024 nodes, 8192 cores). The overall calculation time was 95 days on HOKUSAI BigWave, 346 days on QUEST, and 126 days on RICC. The calculation started on Dec. 30, 2016 and finished on July 9, 2018. The summarization process ended on Oct. 12, 2018.

The whole dataset is available at \url{http://pubchemqc.riken.jp/pm6_dataset.html}. The total size of all files is approximately 800GB. The compressed \texttt{Compounds\_sorted\_20160829.xz} file contains a record for each molecule. Each record is composed of the following fields: CID, molecular weight, InChI representation, isomeric SMILES, molecular compositional formula, electronic charge, and spin number. Another compressed file, \texttt{Compounds\_opt\_failed\_ver1.0.xz}, contains the same information as the above, but for failed compounds. The main data are divided into 4856 files, each of which has a name like \texttt{Compound\_XXXXXXXXX\_XXXXXXXXX.tar.xz}, \\ where \texttt{XXXXXXXXX\_XXXXXXXXX} denotes the range of CIDs. Each o file contains at most 25,000 compounds. Thus, the first file is \texttt{Compound\_000000001\_000025000.tar.xz}, the second file is \texttt{Compound\_000025001\_000050000.tar.xz}, and so forth. 

Each of the main data files contains directories prefixed with CIDs. For example, directory \texttt{000000001} has the following content:
\begin{verbatim}
000000001.20160829.info 000000001.PM6.InChIsame 000000001.PM6.S0.inp 
000000001.PM6.S0.json 000000001.PM6.S0.mulliken 000000001.PM6.anion.inp 
000000001.PM6.anion.json 000000001.PM6.anion.mulliken 000000001.initial.xyz
\end{verbatim}
The \texttt{.info} file contains CID, molecular weight, InChI, isomeric SMILES, charge, and spin number. The \texttt{.initial.xyz} file contains Cartesian coordinates of the atoms of the molecule. The \texttt{.inp} files contain input files, where \texttt{S0} means the ground state of the singlet state. The other symbols, \texttt{D0}, \texttt{T0}, and \texttt{Q0}, denote the ground state of the doublet, triplet, and quartet, respectively. The \texttt{D0} and \texttt{Q0} states for CID 6840 are as follows.
\begin{verbatim}
000006840.20160829.info 000006840.PM6.D0.inp 000006840.PM6.D0.json
000006840.PM6.D0.mulliken 000006840.PM6.InChIsame 000006840.PM6.Q0.inp
000006840.PM6.Q0.json 000006840.PM6.Q0.mulliken 000006840.PM6.cation.inp
000006840.PM6.cation.json 000006840.PM6.cation.mulliken 000006840.initial.xyz
\end{verbatim}
The \texttt{T0} and \texttt{Q0} states are spin flipped states of \texttt{S0} and \texttt{D0}, respectively.

When the original InChI and the calculated InChI at the optimized geometry of the ground state coincided, we added \texttt{.InChIsame} file.  The \texttt{*.cation.*} and \texttt{*.anion.*} files contain the PM6 optimized geometry of cationic and anionic states starting from the \texttt{S0} state or \texttt{D0} state. The JSON files were generated passing the output files of Gaussian09 through cclib.

The \texttt{.mulliken} files contain the Mulliken population of each \texttt{T0}, \texttt{Q0}, \texttt{S0} and \texttt{D0} state. Since cclib canot be used to extract the Mulliken population for an odd number of electronic systems, we used the Mulliken population part from the Gaussian09 log files instead. 

When you extract \texttt{Compound\_000000001\_000025000.tar.xz} file, you will find the same CIDs, 2, 103, and 112 for instance, are missing in the data files.  For instance, CIDs 2 and 112 were omitted since the molecules are charged, while calculations were not performed for CID 103 because the molecular weight, 1156.081 g/mol, exceeds the size limit of 1000g/mol. 

Table~\ref{statistics} lists the statistics of the PM6 geometry optimization on the molecules in PubChem Compound obtained on Aug 29, 2016.  The rows list the corresponding number of compounds and file names containing the detailed data, i.e., the set of PubChem CID, molecular weight, InChI, isomeric SMILES, and molecular formula. For instance, the ``MW less than 1000'' row contains the number of molecules in PubChem Compound whose molecular weight is less than 1000, while the ``Charged molecules'' row lists the number and the file name containing the corresponding detailed data on the charged molecules. The ``No results'' row lists the number of molecules whose PM6 geometry optimization failed for some reason and the\texttt{Compounds\_no\_result\_ver1.0.xz} file contains the detailed data on these instances. The ``InChI (in)valid'' row shows the number of molecules for which the original InChI and calculated InChI (did not) coincide up to the chemical formula and the atom connection of the main layer in the PM6 optimized geometry. The ``Cations'', ``Anions'', and ``Spin flipped'' rows refer to molecules for which we successfully calculated cationic, anionic, spin flipped states starting from the PM6 optimized geometry neutral state. These states are more unstable than neutral molecules and thus are more difficult to calculate. The grand total is the sum of neutral states, cationic states, anionic states, and spin flipped states.

\begin{table}[ht]
\centering
\caption{Statistics of PubChemQC PM6.  We calculated 86,213,135 neutral molecules in total. Among them are 85,197,307 molecules whose original InChI and calculated InChI in PM6 optimized geometry coincided up to the chemical formula and the atom connection of the main layer. The list of molecules with CID, molecular weight, InChI, SMILES are available at \texttt{http://pubchemqc.riken.jp/pm6\_dataset.html}}
 \begin{tabular}{lrr}
 \hline \hline
 Results & Molecule count & Filename \\ \hline
 PubChem Compound & 91,679,247 & \texttt{Compounds\_all\_sortedbymw\_ver1.0.xz} \\ 
 MW less than 1000 & 91,074,917 & - \\ 
 Charged molecules & 2,188,881 & \texttt{Compounds\_charged\_ver1.0.xz} \\ 
 No results & 2,672,901 & \texttt{Compounds\_no\_result\_ver1.0.xz} \\ 
 Calculated & 86,213,135 & - \\ 
 InChI valid & 85,197,307 & \texttt{Compounds\_inchi\_valid\_ver1.0.xz} \\ 
 InChI invalid & 1,015,828 & \texttt{Compounds\_inchi\_invalid\_ver1.0.xz} \\ 
 Cations & 51,555,911 & \texttt{Compounds\_cation\_ver1.0.xz} \\
 Anions & 45,581,750 & \texttt{Compounds\_anion\_ver1.0.xz} \\ 
 Spin flipped & 37,839,619 & \texttt{Compounds\_spinflip\_ver1.0.xz}\\
 Grand total & 221,190,415 & - \\
 \hline \hline
 \end{tabular}
 \label{statistics}
\end{table}

Table~\ref{statistics2} lists the elements appearing in the compounds. For C6H2, we count C as one and H as one.  An entry in the ``Success'' column means the geometry optimization in the neutral state was successful. ``InChI valid'' means the original InChI and the calculated InChI coincide up to the chemical formula and the atom connection of the main layer in the PM6 optimized geometry. ``Failed'' means the PM6 geometry optimization failed. ``Total'' means the number of molecules in the PM6 calculation. The calculations for many non-metal elements were successful. In contrast, the calculations failed for all of the lanthanoids, except La, Gd, and Lu. Moreover, the calculations failed for many metals. For example, while 12,691 molecules in PubChem Compounds contain Fe, the calculation succeeded for only 1,363 of those molecules. Determining why the calculations failed for metals will be a future task. 

\begin{longtable}{lrrrr}
\caption{Statistics of PubChemQC PM6 calculation by element. Note that CH4O2 is counted as one carbon, one hydrogen, and one oxygen. An entry in the “Success” column means the geometry optimization of PM6 in the neutral state was successful, ``InChI valid'' means the original InChI and the calculated InChI coincide up to the chemical formula and the atom connection of the main layer in the PM6 optimized geometry. ``Failed'' means the PM6 geometry optimization failed. ``Total'' means the number of molecules in the PM6 calculation. }
\label{statistics2}
 \\
 \hline \hline
 Element & Success & InChI valid & Failed & Total \\ \hline
 \endfirsthead
 \multicolumn{5}{r}{Continued} \\ \hline 
 Element & Success & InChI valid & Failed & Total \\ \hline
 \endhead
 \hline
 \multicolumn{5}{r}{Continue} \\ 
 \endfoot
 \hline
 \endlastfoot
 H & 86,190,789 & 85,177,145 & 2,659,904 & 88,850,693 \\
 He & 38 & 38 & 17 & 55 \\
 Li & 25,535 & 20,935 & 5,080 & 30,615 \\
 Be & 241 & 115 & 287 & 528 \\
 B & 233,224 & 214,868 & 36,902 & 270,126 \\
 C & 86,203,117 & 85,189,423 & 2,653,459 & 88,856,576 \\
 N & 79,618,703 & 78,739,413 & 1,993,733 & 81,612,436 \\
 O & 77,016,235 & 76,134,745 & 2,388,262 & 79,404,497 \\
 F & 16,150,043 & 15,997,142 & 394,265 & 16,544,308 \\
 Ne & 15 & 15 & 7 & 22 \\
 Na & 121,065 & 57,346 & 23,232 & 144,297 \\
 Mg & 11,553 & 8,287 & 3,732 & 15,285 \\
 Al & 8,275 & 6,561 & 8,405 & 16,680 \\
 Si & 603,361 & 580,547 & 197,755 & 801,116 \\
 P & 715,639 & 607,563 & 71,616 & 787,255 \\
 S & 27,182,250 & 26,644,802 & 621,710 & 27,803,960 \\
 Cl & 15,983,905 & 15,775,154 & 493,800 & 16,477,705 \\
 Ar & 154 & 150 & 120 & 274 \\
 K & 36,393 & 29,666 & 8,633 & 45,026 \\
 Ca & 7,063 & 4,859 & 5,045 & 12,108 \\
 Sc & 47 & 36 & 485 & 532 \\
 Ti & 2,310 & 1,275 & 17,308 & 19,618 \\
 V & 92 & 66 & 4,054 & 4,146 \\
 Cr & 691 & 399 & 5,337 & 6,028 \\
 Mn & 125 & 104 & 4,679 & 4,804 \\
 Fe & 2,746 & 1,363 & 9,945 & 12,691 \\
 Co & 772 & 583 & 7,302 & 8,074 \\
 Ni & 2,092 & 1,123 & 9,965 & 12,057 \\
 Cu & 5,777 & 3,731 & 9,562 & 15,339 \\
 Zn & 12,335 & 7,996 & 6,581 & 18,916 \\
 Ga & 1,150 & 1,006 & 1,896 & 3,046 \\
 Ge & 8,093 & 7,230 & 2,666 & 10,759 \\
 As & 10,667 & 6,753 & 6,329 & 16,996 \\
 Se & 46,401 & 39,790 & 3,682 & 50,083 \\
 Br & 6,424,739 & 6,339,585 & 156,506 & 6,581,245 \\ 
 Kr & 30 & 30 & 18 & 48 \\
 Rb & 2,353 & 1,957 & 1,066 & 3,419 \\
 Sr & 659 & 498 & 695 & 1,354 \\
 Y & 428 & 232 & 3,715 & 4,143 \\
 Zr & 3,533 & 983 & 15,957 & 19,490 \\
 Nb & 82 & 70 & 710 & 792 \\
 Mo & 372 & 263 & 2,425 & 2,797 \\
 Tc & 109 & 43 & 1,107 & 1,216 \\
 Ru & 1,386 & 829 & 6,098 & 7,484 \\
 Rh & 325 & 215 & 1,930 & 2,255 \\
 Pd & 3,413 & 1,975 & 4,288 & 7,701 \\
 Ag & 3,169 & 2,609 & 1,371 & 4,540 \\
 Cd & 928 & 683 & 965 & 1,893 \\
 In & 939 & 829 & 1,381 & 2,320 \\
 Sn & 31,405 & 29,174 & 18,900 & 50,305 \\
 Sb & 2,673 & 1,155 & 2,472 & 5,145 \\
 Te & 6,076 & 4,417 & 875 & 6,951 \\
 I & 1,561,730 & 1,475,887 & 201,664 & 1,763,394 \\
 Xe & 53 & 45 & 104 & 157 \\
 Cs & 972 & 500 & 346 & 1,318 \\
 Ba & 1,727 & 664 & 1,941 & 3,668 \\
 La & 99 & 63 & 774 & 873 \\
 Ce & 0 & 0 & 959 & 959 \\
 Ba & 1,727 & 664 & 1,941 & 3,668 \\
 La & 99 & 63 & 774 & 873 \\
 Ce & 0 & 0 & 959 & 959 \\
 Pr & 0 & 0 & 569 & 569 \\
 Nd & 0 & 0 & 797 & 797 \\
 Pm & 0 & 0 & 38 & 38 \\
 Sm & 0 & 0 & 566 & 566 \\
 Eu & 0 & 0 & 845 & 845 \\
 Gd & 82 & 57 & 937 & 1,019 \\
 Tb & 0 & 0 & 410 & 410 \\
 Dy & 0 & 0 & 309 & 309 \\
 Ho & 0 & 0 & 255 & 255 \\
 Er & 0 & 0 & 389 & 389 \\
 Tm & 0 & 0 & 234 & 234 \\
 Yb & 0 & 0 & 503 & 503 \\
 Lu & 37 & 29 & 238 & 275 \\
 Hf & 538 & 144 & 2,242 & 2,780 \\
 Ta & 89 & 80 & 626 & 715 \\
 W & 545 & 347 & 5,308 & 5,853 \\
 Re & 69 & 58 & 961 & 1,030 \\
 Os & 83 & 72 & 938 & 1,021 \\
 Ir & 572 & 313 & 3,219 & 3,791 \\
 Pt & 3,836 & 1,610 & 10,870 & 14,706 \\
 Au & 1,322 & 883 & 1,569 & 2,891 \\
 Hg & 3,110 & 1,946 & 4,717 & 7,827 \\
 Tl & 701 & 478 & 565 & 1,266 \\
 Pb & 2,602 & 2,235 & 1,888 & 4,490 \\
 Bi & 1,364 & 1,104 & 1,965 & 3,329 \\
 \hline \hline
\end{longtable}
 
\section{5. Discussion}
\label{sec:discuss}
We successfully calculated 86,213,135 compounds, covering 94.0\% of those in the PubChem Compound dataset. Moreover, the original and calculated InChIs coincided up to the chemical formula and the atom connection of the main layer for 92.9\% of compounds in the PubChem Compound dataset. Note that when the calculations failed, we recalculated several times, since Open~Babel generates a slightly different initial geometry guess for each execution. We also calculated cationic, anionic, and spin flipped states starting from the PM6 optimized geometry. These states are unstable, and hence, coverage is low.

We can learn a lot from the remaining 7.1\% of compounds for which the original and calculated InChIs did not coincide or calculations failed.  The following examples illustrate problems in the original data, calculation methods, and molecular encoding methods that led to failures.

We found duplicates in the original PubChem Compound database. For example, CID 5362549 and CID 5460654 are the same hydrogen atom, and CID 5487799 and CID 139073 are the same beryllium hydride. We did not look for duplicates thoroughly, but we believe that there are more exact duplicates. A more problematic example is the boron hydrate anion (CID 15398067 and CID 54713148). The difference between the canonical SMILESs of these two molecules is in the assignment of the formal charge: \texttt{[H-].[B]} and \texttt{[BH-]}. Consequently, the InChIs are also different. However, this does not mean they should be distinguishable; chemically, they are the same species. A similar problem exists for CID 18186120 and CID 961: their SMILES are \texttt{[H+].[O-2]} and \texttt{[OH-]}, respectively.

Another problem is when the InChIs are the same, but the SMILESs are different. For example, CID 14917 and CID 16211014 both have the same InChI \texttt{InChI=1S/FH/h1H}, while the SMILESs are \texttt{F} and \texttt{[H+].[F-]}, respectively. These examples clearly show that InChI and SMILES are not compatible in principle. Moreover, there is no rigorous correspondence between the atomic coordinates and SMILES or InChI. We allow duplicates and ambiguity to some extent. 

We performed the same calculation procedure for salts and mixtures. These systems contain ``\texttt{.}'' in the SMILES representation (for example, CID 16741201). Many of the mixtures are not suitable for calculations. However, it is a not trivial problem to distinguish them from molecules. For example, CID 88524581 is a metal complex, not a mixture, while CID 24670 is a mixture of two compounds. A more complicated case is CID 5351148, which is a hydrochloride salt that looks to be stable as a hydrochloride solution.

We did not consider isotopes; we used the most stable substitute instead. We treated D as H (e.g., CID 5460634), 13C as 12C (e.g., CID 101192347), and 15N as N. Moreover, we did not calculate charged ions. For example, CID 1038 is a hydrogen ion ($\rm H^+$), and CID 166653 is a hydride ($\rm H^-$). We regarded such systems as ionized species of CID 5362549 (hydrogen). On the other hand, we took isomers into account when different CIDs had been assigned to two different isomers. We were not interested in isomers or conformers of stearic acid (CID 5281). Open~Babel chooses an isomer as an initial guess.

There are some molecules whose original InChI and calculated InChI are different. The following two examples illustrate why we only checked the identity of the chemical formula and atom connection of the main layer.

\begin{itemize}
\item\textbf{Difference between strictnesses of stereochemical layers.}\\
For CID 179, the original InChI in PubChem Compound is
\begin{verbatim}
InChI=1S/C4H8O2/c1-3(5)4(2)6/h3,5H,1-2H3
\end{verbatim}
whereas the calculated InChI is
\begin{verbatim}
InChI=1S/C4H8O2/c1-3(5)4(2)6/h3,5H,1-2H3/t3-/m0/s1
\end{verbatim}

The calculated InChI treats the stereochemical layer more strictly. Figure~\ref{CID:179} shows there is no significant difference between the ball-and-stick models.
\begin{figure}
\caption{Ball-and-stick models for CID 179. Left: initially generated by Open~Babel. Right: PM6 calculated molecules. These molecules are almost the same, but the original InChI (\texttt{InChI=1S/C4H8O2/c1-3(5)4(2)6/h3,5H,1-2H3} and calculated InChI (\texttt{InChI=1S/C4H8O2/c1-3(5)4(2)6/h3,5H,1-2H3/t3-/m0/s1}) are different because the stereo chemistry is more strictly treated in the calculated InChI .}
\label{CID:179}
 \begin{tabular}{cc}
 \includegraphics[width=5cm]{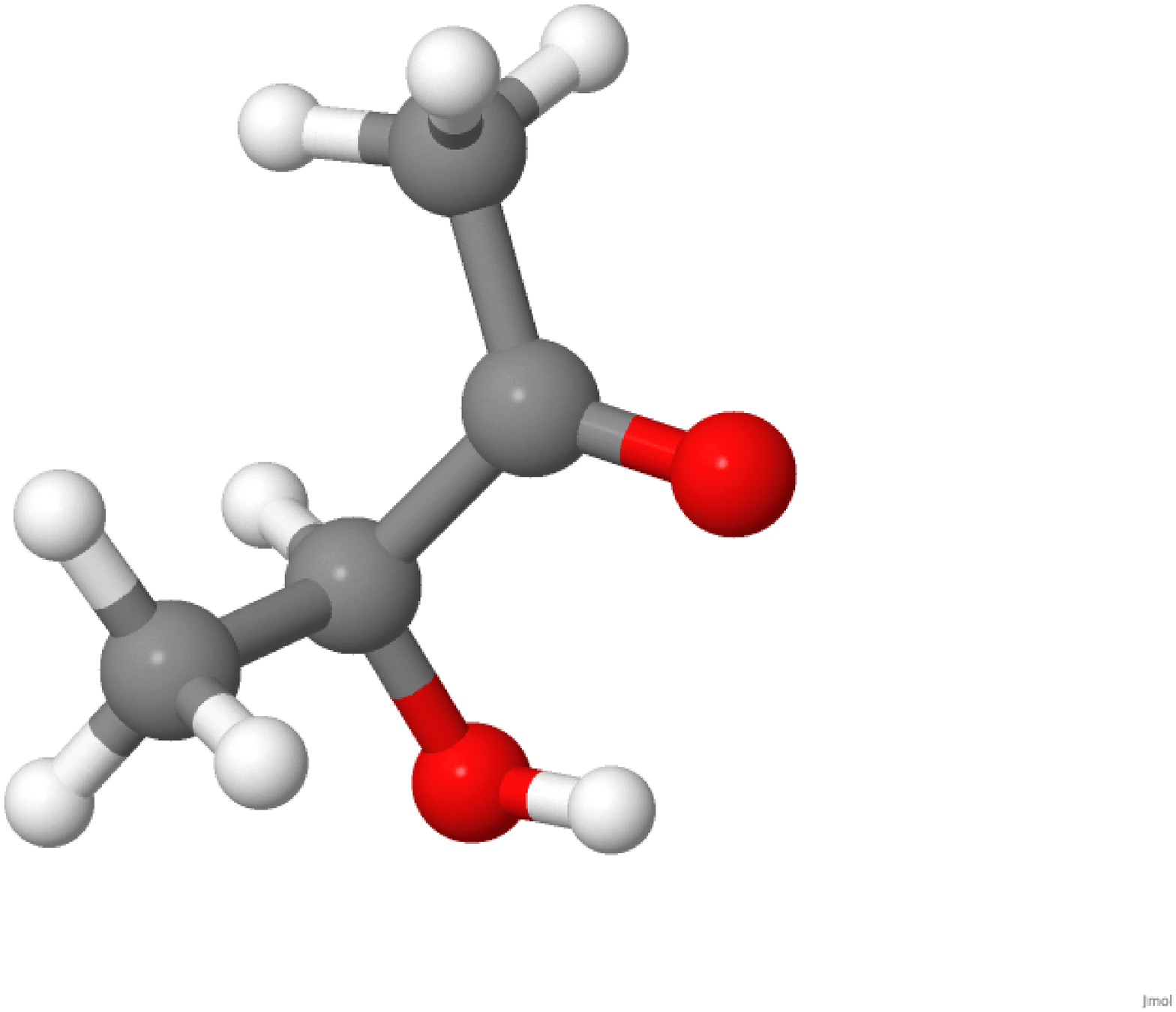} & \includegraphics[width=5cm]{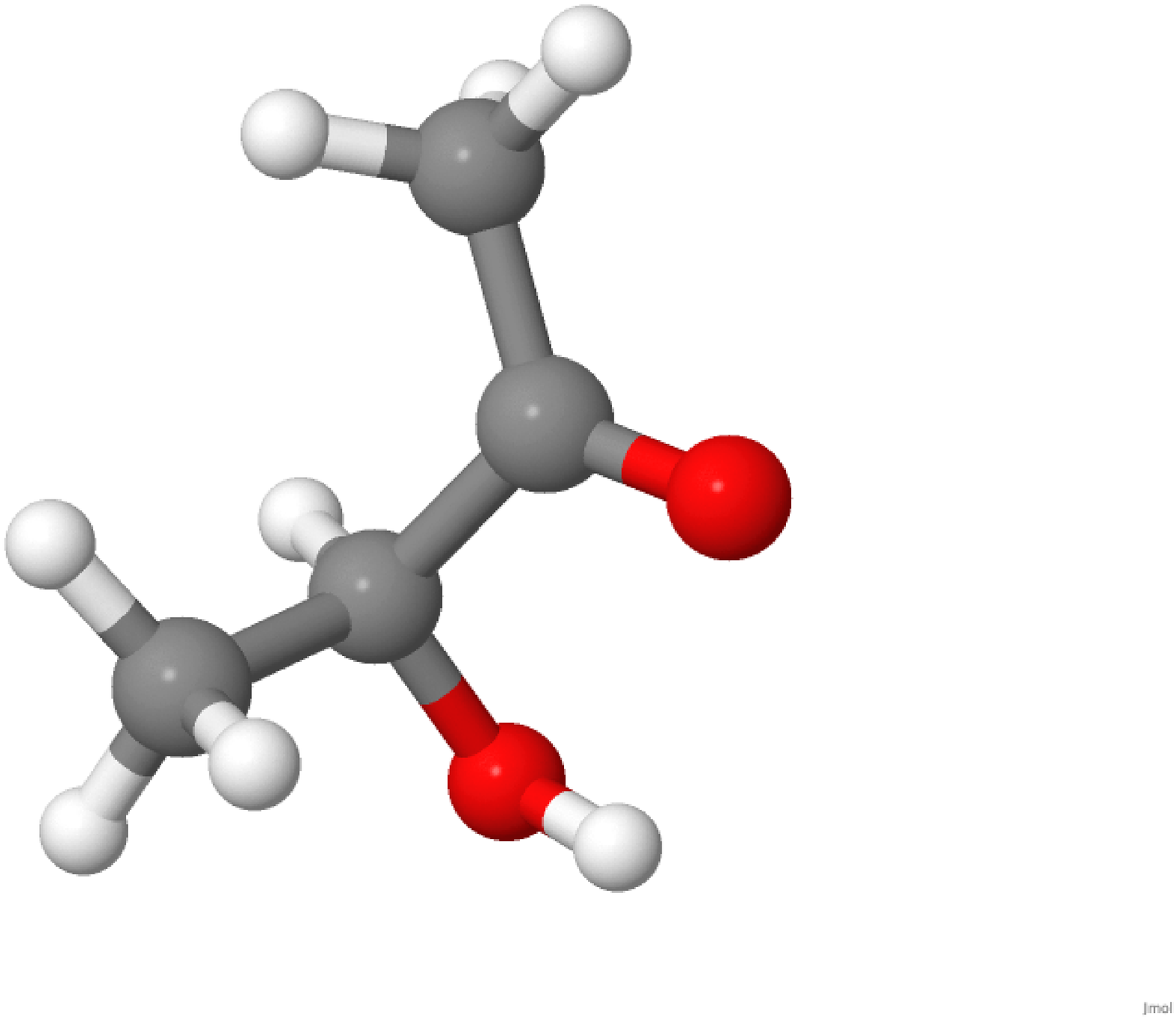} \\
 \end{tabular}
\end{figure}

\item\textbf{Difference between strictnesses of assignment of implicit hydrogen or treatment of tautomers.}\\
For CID 5987, the original InChI is
\begin{verbatim}
InChI=1S/H3NO3S/c1-5(2,3)4/h(H3,1,2,3,4)
\end{verbatim}
and the calculated InChI is
\begin{verbatim}
InChI=1S/H3NO3S/c1-5(2,3)4/h2H,1H2
\end{verbatim}
Figure~\ref{CID:5987} shows the ball-and-stick models. The apparent discrepancy arises because quantum chemical calculations make floating hydrogen explicit.
\begin{figure}
\caption{Ball-and-stick models for CID 5987. Left: initially generated by Open~Babel ({\tt InChI=1S/H3NO3S/c1-5(2,3)4/h(H3,1,2,3,4)}). Right: molecules calculated by PM6 ({\tt InChI=1S/H3NO3S/c1-5(2,3)4/h2H,1H2}). The hydrogens are originally floating, but are explicit in the calculated InChI. There is no general solution to this problem.}\label{CID:5987}
 \begin{tabular}{cc}
 \includegraphics[width=5cm]{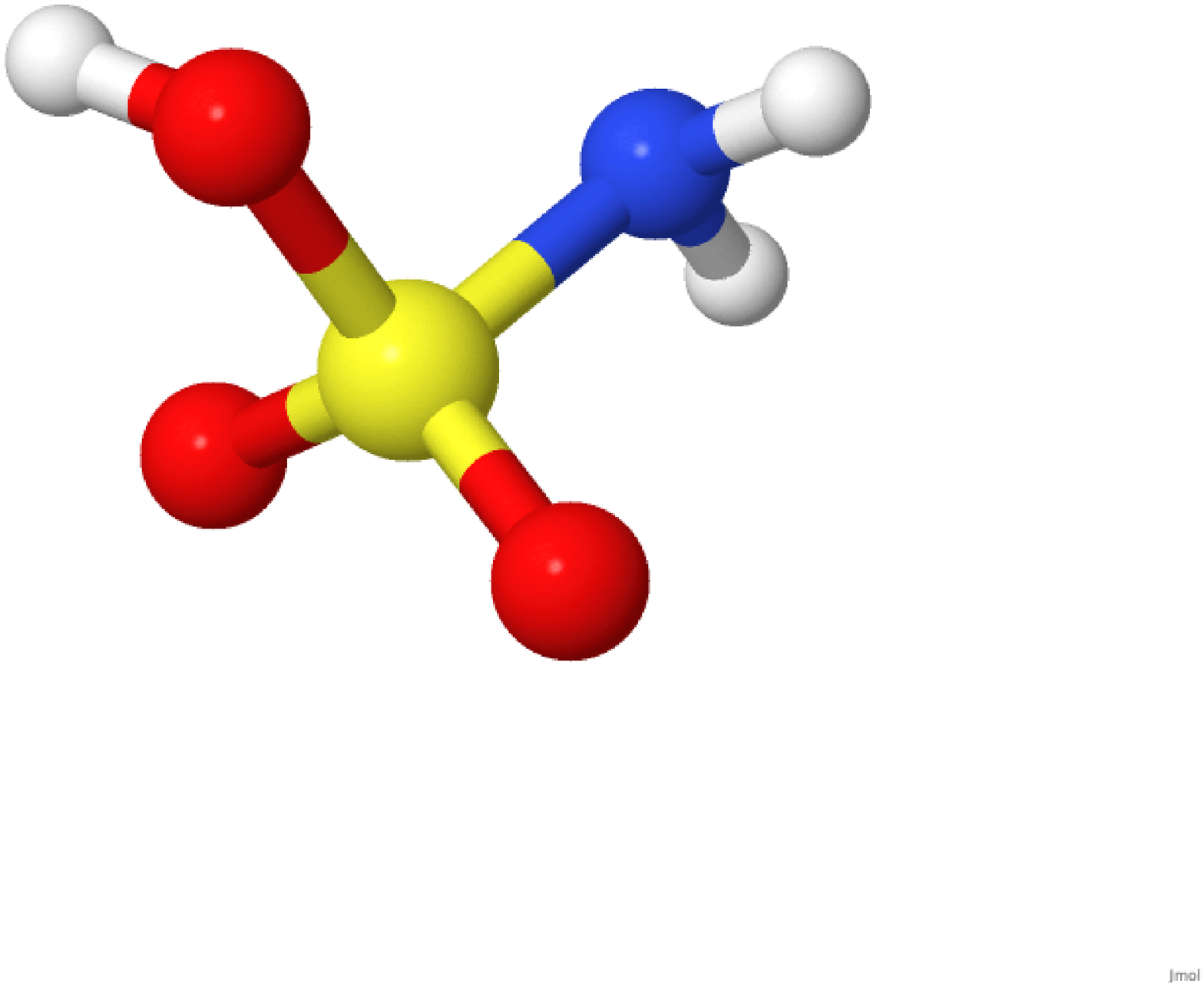} & \includegraphics[width=5cm]{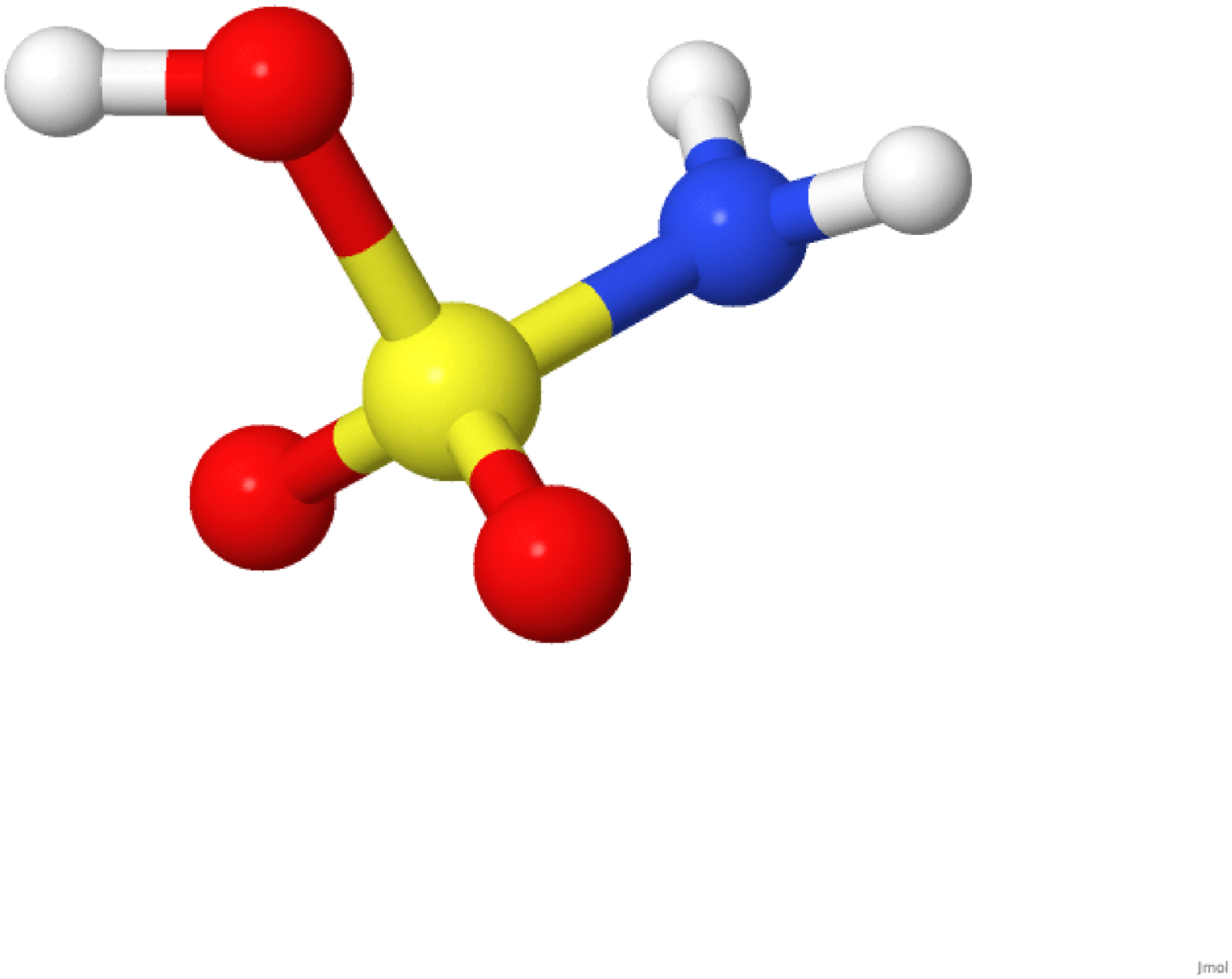} \\
 \end{tabular}
\end{figure}
The above examples are cases in which Open~Babel always specifies the hydrogens or stereochemistry in the calculation of InChI when the original InChI contains some ambiguity. Consequently, the two InChIs contradict each other. Unfortunately, there is no general solution to this problem.

The next example shows a case where our approach does not work for solids or solutions in principle.
\item\textbf{Different interpretations of the hydrogen atom.}\\
Tetramethylammonium hydroxide, CID 60966, has the chemical formula $\rm [(CH_3)_4 N]^+][OH]^-$. Tetramethylammonium hydroxide pentahydrate is in a stable solid state or dissolved in water or methanol solution. The InChI given by the initial geometry guess and the InChI given by PM6 optimized geometry are as follows:
\begin{verbatim}
InChI=1S/C4H12N.H2O/c1-5(2,3)4;/h1-4H3;1H2/q+1;/p-1
InChI=1S/C4H12N.HO/c1-5(2,3)4;/h1-4H3;1H
InChI=1S/C4H11N.H2O/c1-5(2,3)4;/h1H2,2-4H3;1H2
\end{verbatim}

The three InChI representations are different. The second InChI keeps the original intention. (See Figure~\ref{CID:60966}.) The PM6 geometry optimization was performed in the gas state, so the hydrate ion took the hydrogen and was stabilized as water. Therefore, we obtained a different representation from the original InChI.  Fortunately, both the original and the PM6 optimized InChIs are the same. Hence, the calculation was successful in this case.
\begin{figure}
\caption{Ball-and-stick models for CID 60966. Left: initially generated by Open~Babel. Right: molecules calculated by PM6 geometry optimization. In the PM6 optimization, a hydrogen moved to an oxygen. We performed the whole calculation in the gas phase. Therefore, the InChI of the PM6 calculation is different from the original one.}\label{CID:60966}
\begin{tabular}{cc}
\includegraphics[width=5cm]{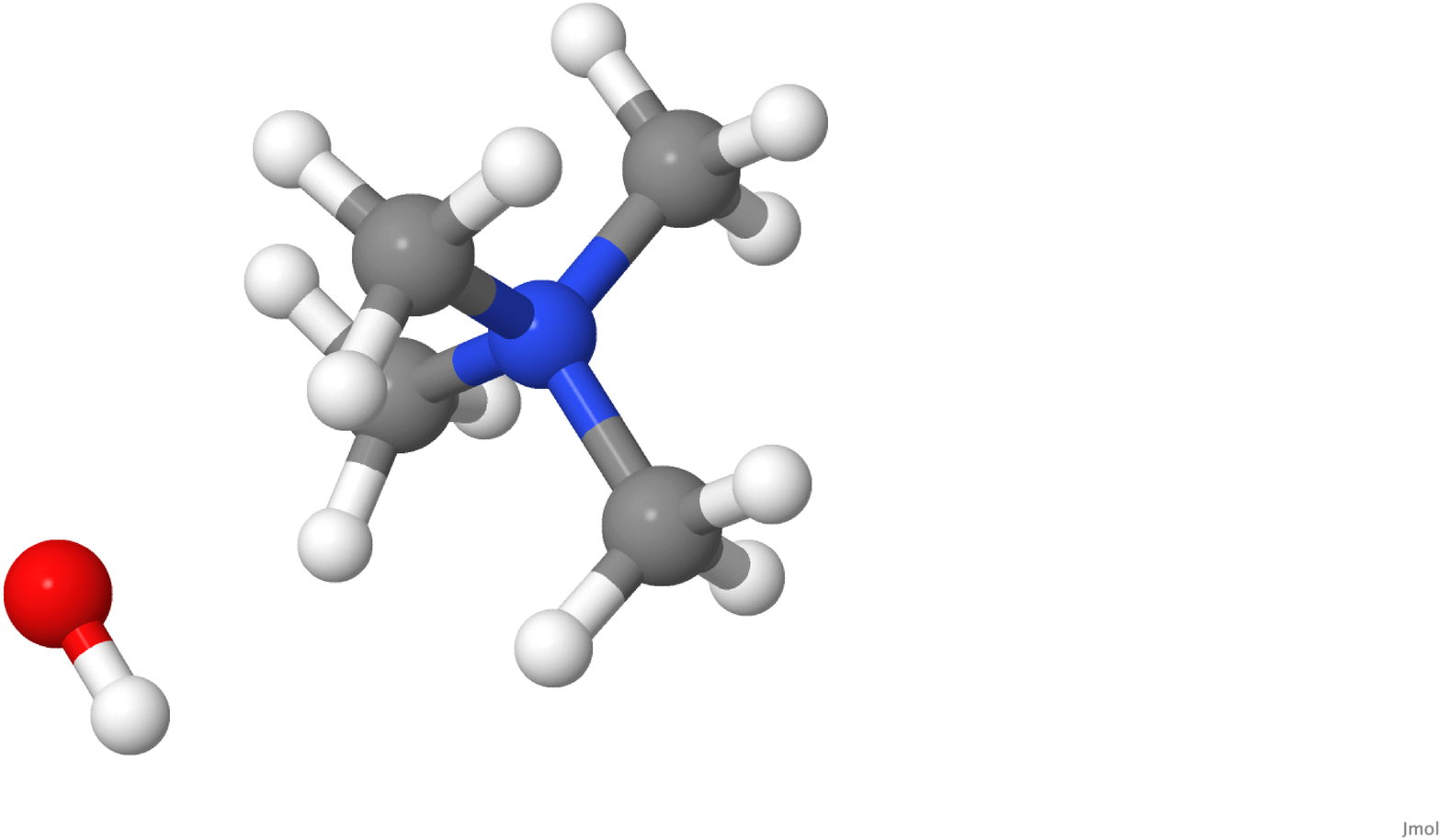} & \includegraphics[width=5cm]{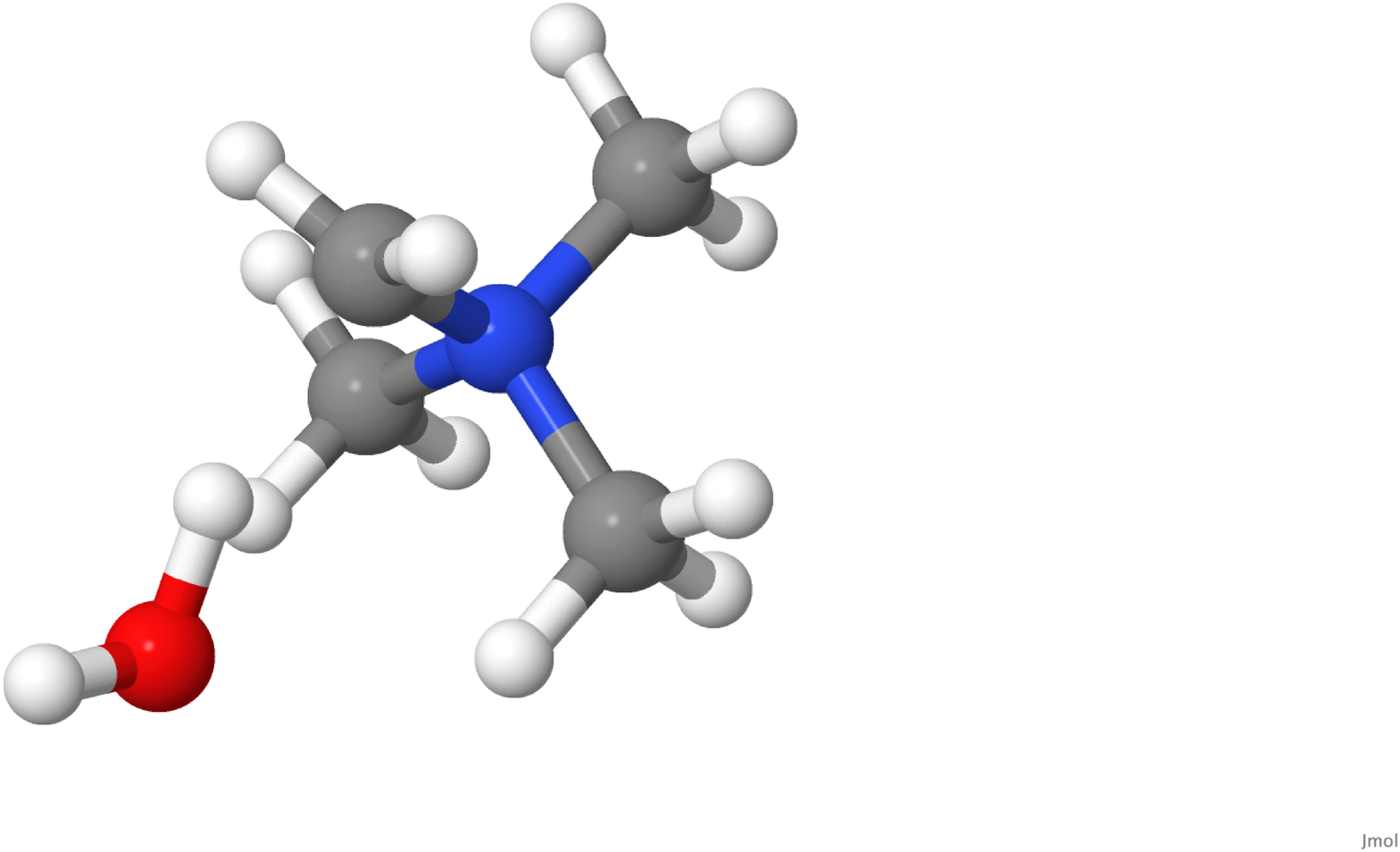} \\
\end{tabular}
\end{figure}

\item\textbf{Difference between the original geometry and the PM6 optimized geometry.}\\
For CID 53628168, the original InChI and the InChI calculated from the initial geometry guess are the same:
\begin{verbatim}
InChI=1S/C4H7N/c1-4(2)3-5/h5H,1-3H2
\end{verbatim}
However, InChI in the PM6 optimized geometry is:
\begin{verbatim}
InChI=1S/C4H7N/c1-4-2-5-3-4/h5H,1-3H2
\end{verbatim}
In this case, PM6 geometry optimization has changed the molecule from the original InChI representation of CID 53628168 (see Figure~\ref{CID:53628168}); a ring is formed.  
\begin{figure}
\caption{Ball-and-stick models for CID 53628168. Left: initially generated by Open~Babel. Middle: initial chemical formula. Right: molecules calculated by PM6 geometry optimization. The original compound looks very unstable. The PM6 calculation stabilized the original compound by forming a ring. This substantially changed the InChI.} \label{CID:53628168}
 \begin{tabular}{ccc}
 \includegraphics[width=5cm]{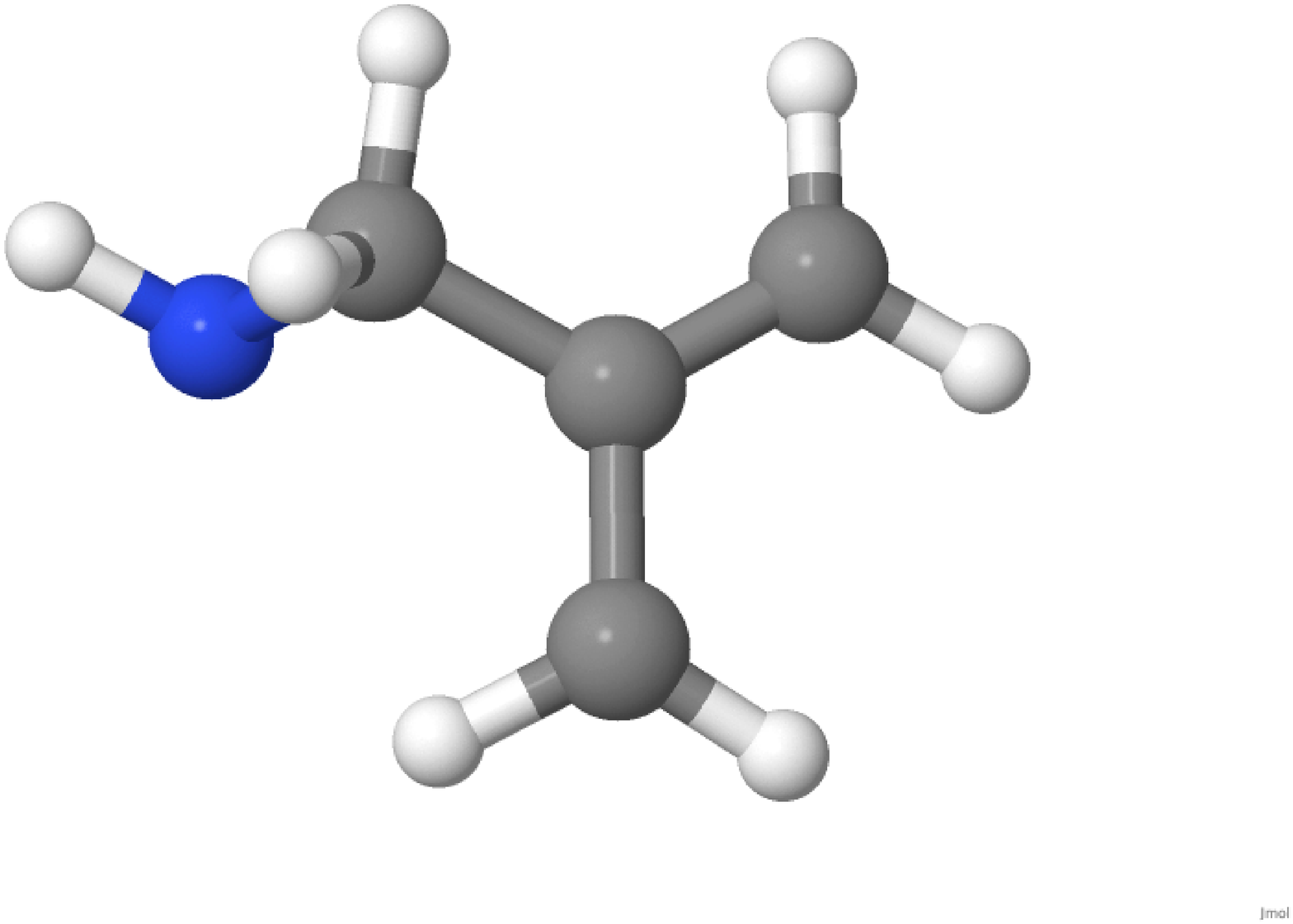} &
 \includegraphics[width=5cm]{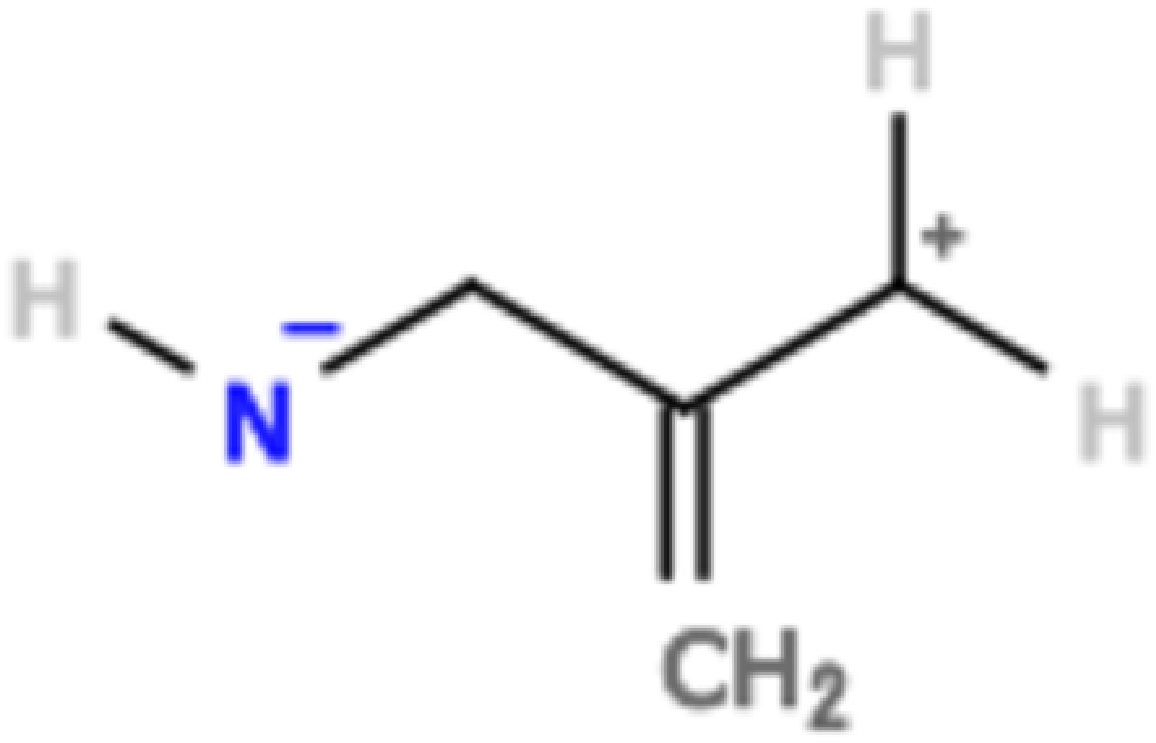} & 
 \includegraphics[width=5cm]{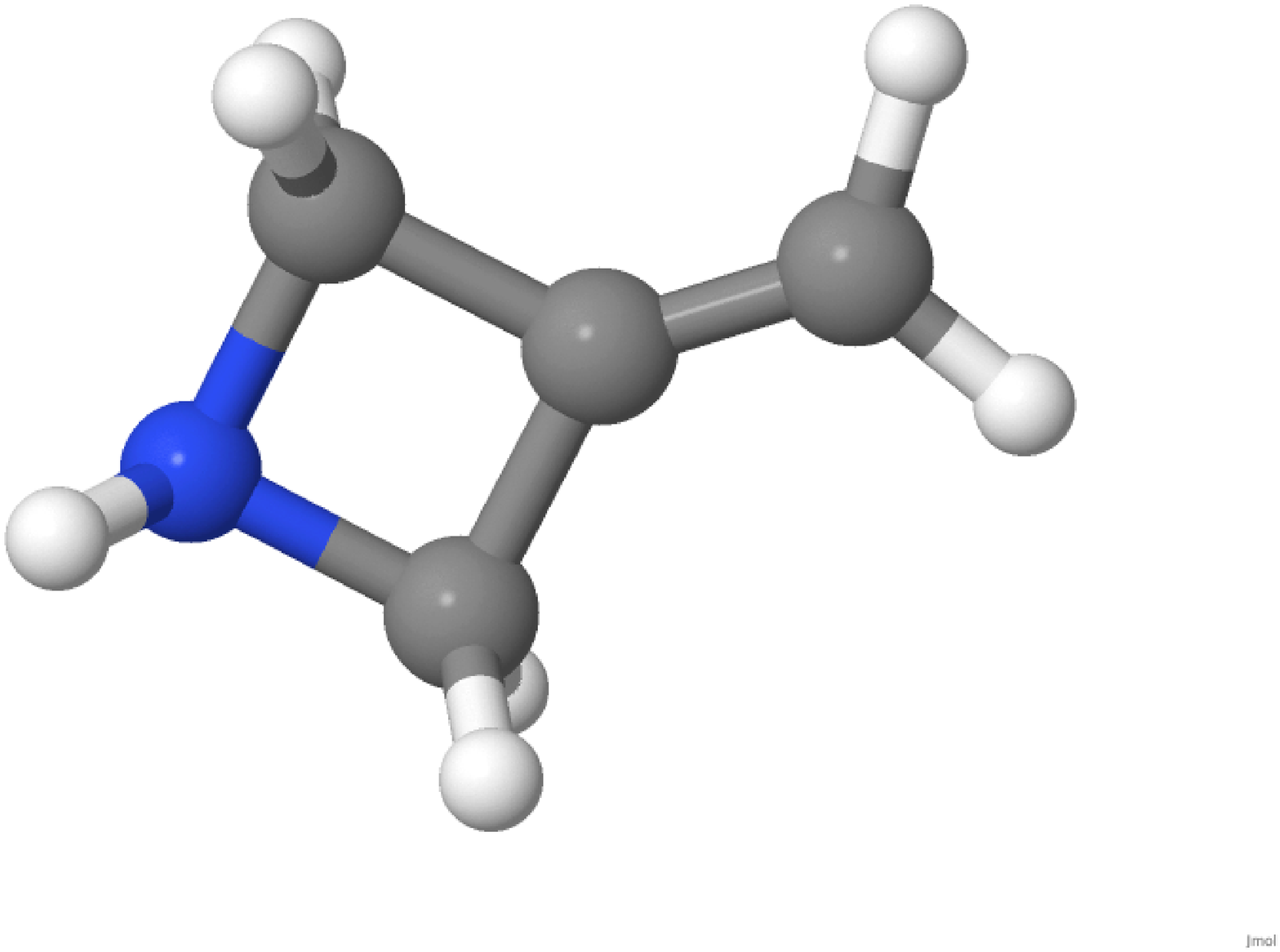} \\
 \end{tabular}
 \end{figure}
As the molecule was modified during the geometry optimization, it may be very unstable. Thus, it would likely form a ring. A similar situation can be seen for CID 53629728 (Pyridazinediyl), where the original and the InChI calculated from the initial geometry guess were the same \texttt{InChI=1S/C4N2/c1-2-4-6-5-3-1},  having a hexagonal shape. However, the InChI given by the PM6 optimized geometry is \texttt{InChI=1S/C4N2/c5-3-1-2-4-6}; it has a linear form.
\begin{figure}
\caption{Ball-and-stick models for CID 53629728. Left: initially generated by Open~Babel. Middle: initial chemical formula. Right: molecules calculated by PM6. The molecular shape of the PM6 calculation is totally different from the shapes of the molecules initially generated by Open~Babel and the chemical formula.} \label{CID:053629728}
\begin{tabular}{ccc}
 \includegraphics[width=5cm]{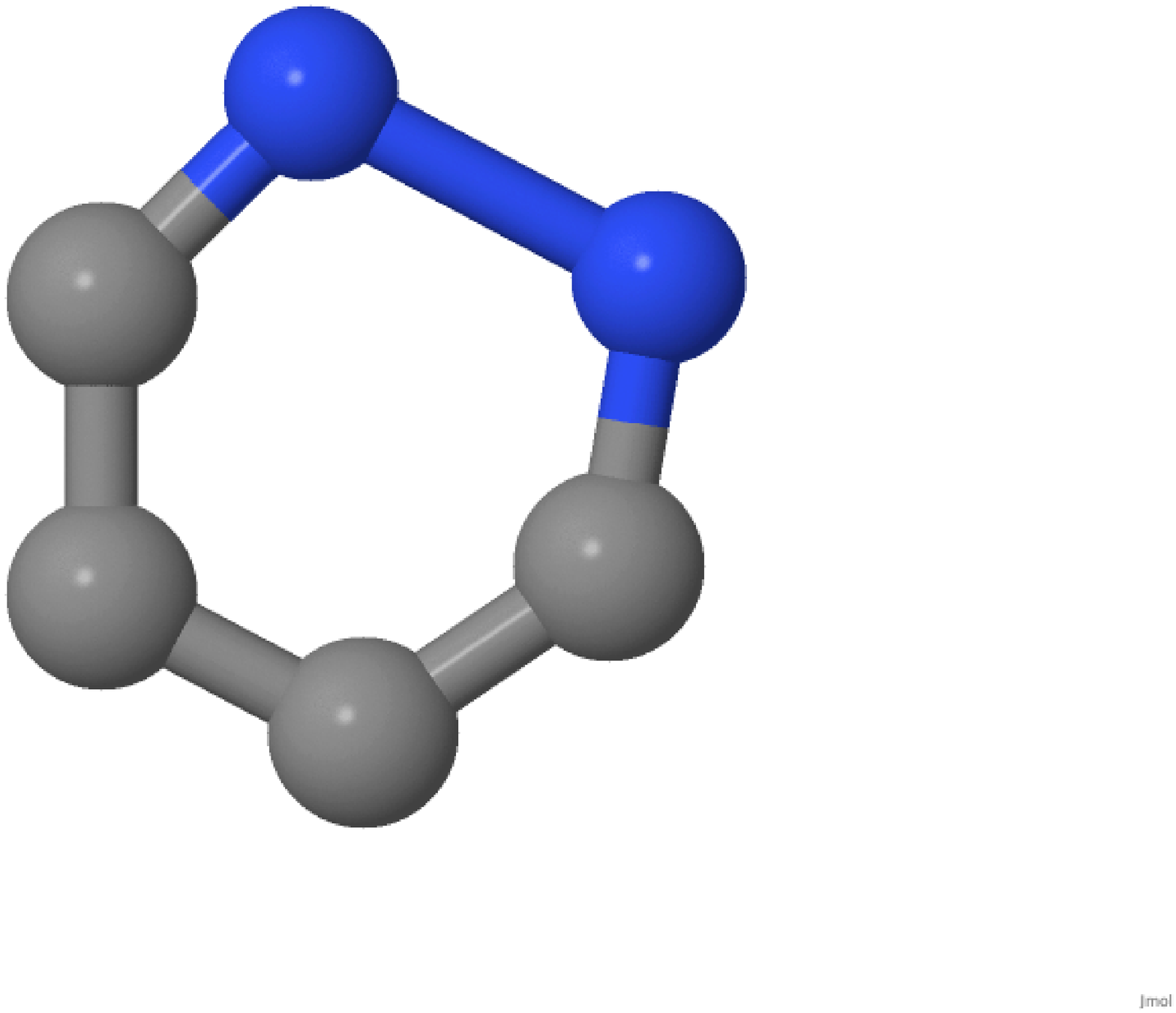} &
 \includegraphics[width=5cm]{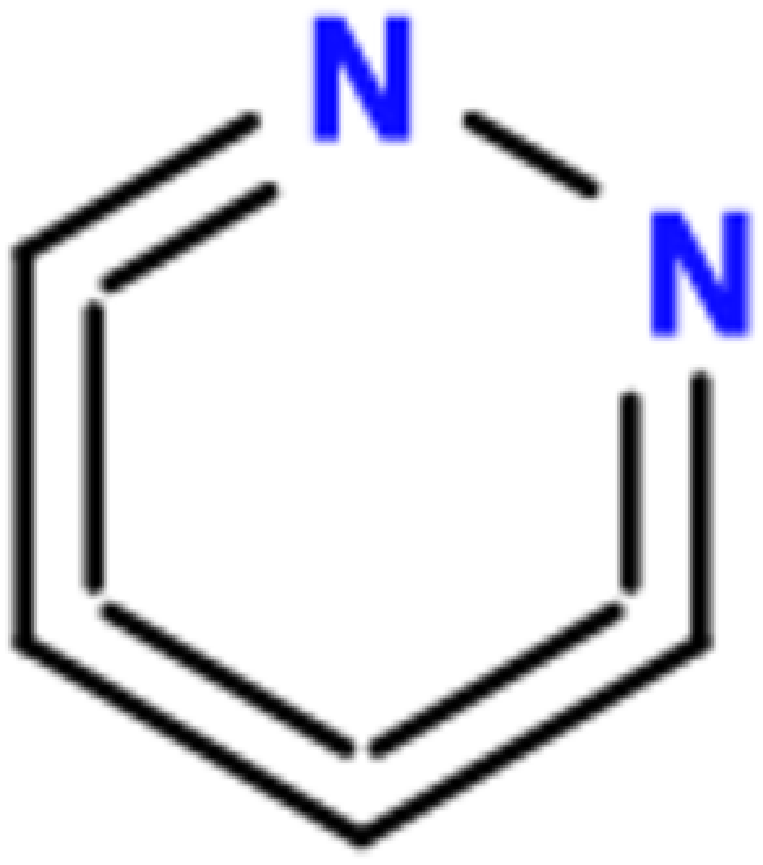} & 
 \includegraphics[width=5cm]{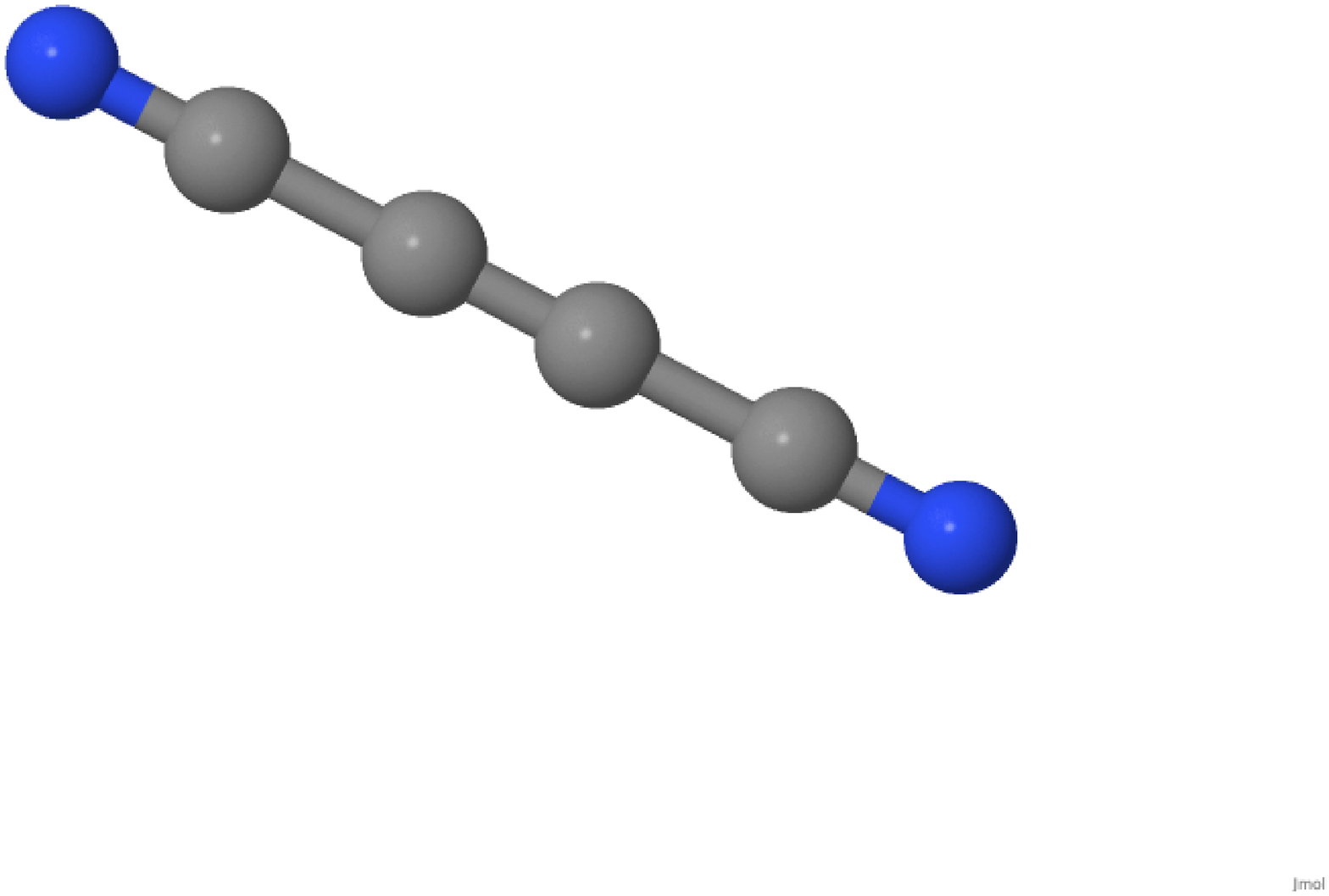} \\
\end{tabular}
\end{figure}
For CID 59269024, the original InChI and the InChI calculated from the initial geometry guess are different:
\begin{verbatim}
InChI=1S/C5H4N/c1-5-3-2-4-6-5/h1-4H/q-1/p+1
\end{verbatim}

The initial geometry guess by Open~Babel is as follows:
\begin{verbatim}
InChI=1S/C5H5N/c1-5-3-2-4-6-5/h1-4,6H
\end{verbatim}

From a quantum chemical point of view, the original InChI might not be an appropriate representation; the formal charges and formal protons do not make sense. Additionally, the molecular formulae are different: the original one is \texttt{C5H4N}, and the calculated one is \texttt{C5H5N}. Depending on the situation, a discrepancy such as this one can be vanished by forbidding the formal proton and electron layer to InChI.
\begin{figure}
\caption{Ball-and-stick models for CID 59269024. Left: initially generated by Open~Babel. Middle: initial chemical formula. Right: molecules calculated by PM6 geometry optimization. These three molecules represent the same molecules, but their InChIs are different because the formal charges are different.  } \label{CID:059269024}
\begin{tabular}{ccc}
 \includegraphics[width=5cm]{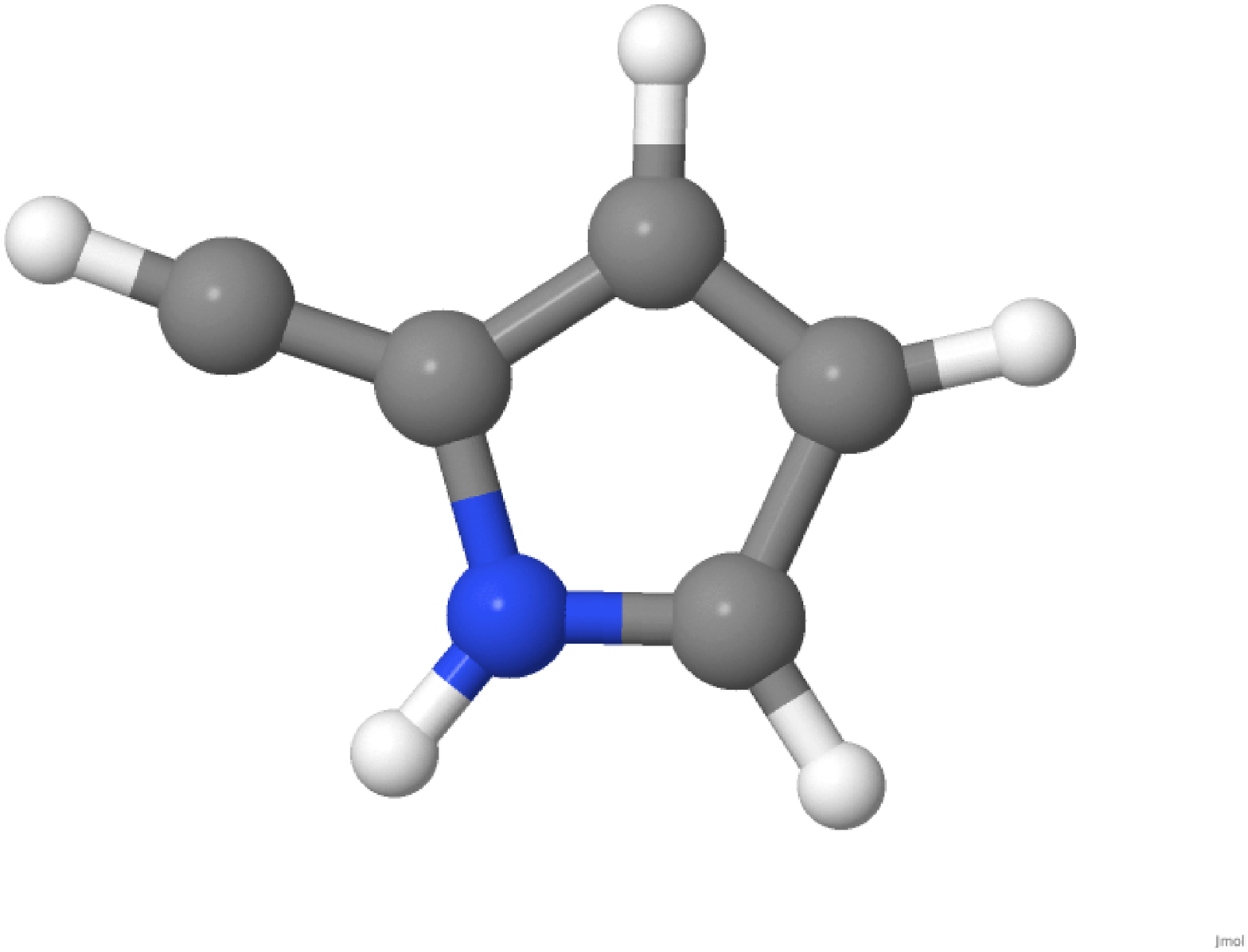} &
 \includegraphics[width=5cm]{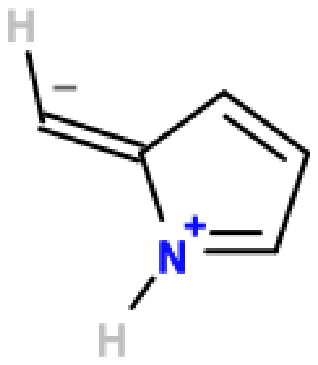} & 
 \includegraphics[width=5cm]{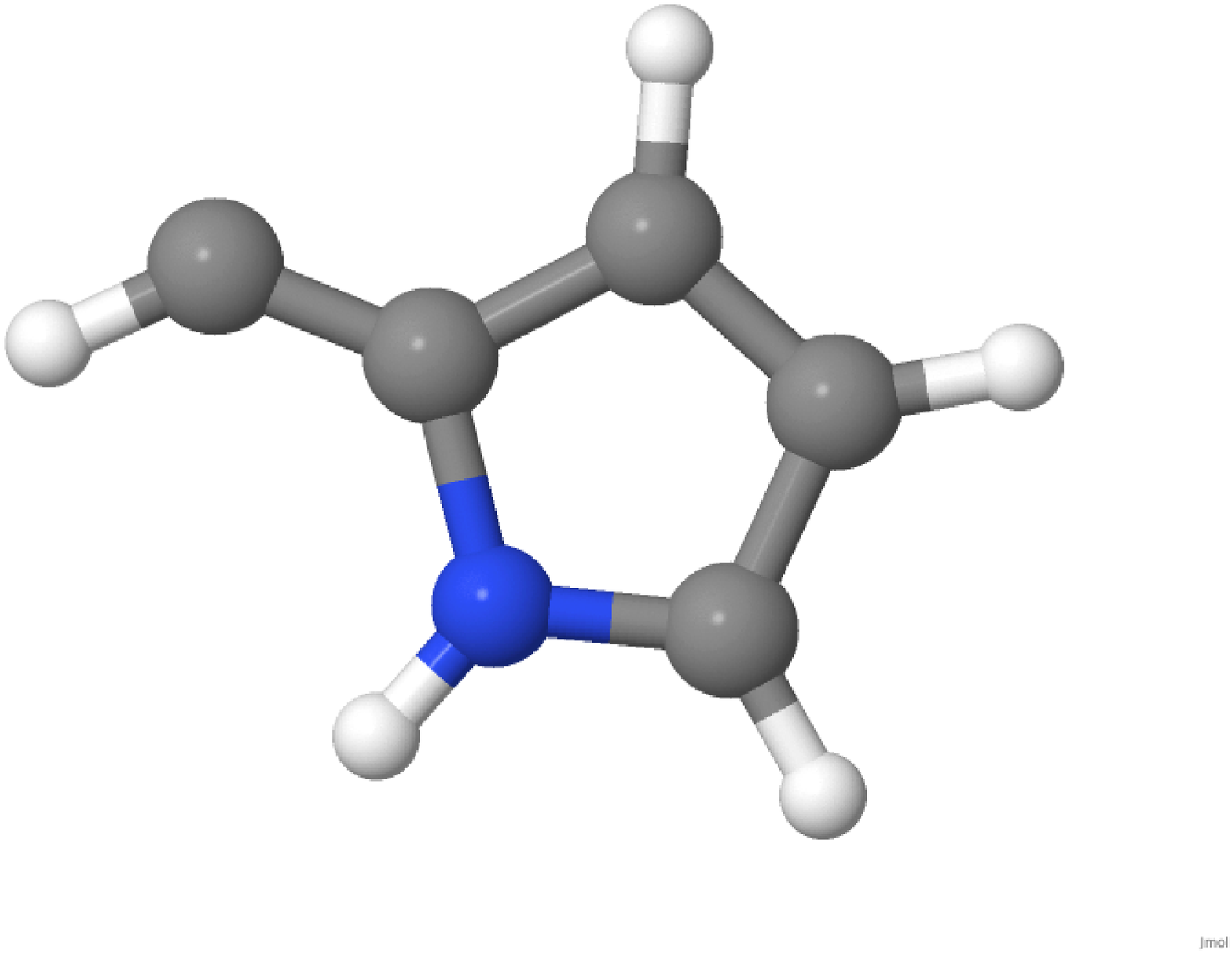} \\
\end{tabular}
\end{figure}

Another complicated example is CID 20149525. Here, the initial InChI and the InChI calculated from the initial geometry guess are the same:
\begin{verbatim}
InChI=1S/C4H6.Si/c1-3-4-2;/h3-4H,1-2H2;
\end{verbatim}
However, the InChI in the PM6 optimized geometry is represented as follows.
\begin{verbatim}
InChI=1S/C4H6Si/c1-2-4-3-5-4/h2,4H,1,3H2/t4-/m1/s1
\end{verbatim}
It differs from the original InChI. In this case, Si is not bonded in the original structure, but Open~Babel considered that Si is bonded to C-C in the optimized structure. The distances between Si and C are 0.191 nm and 0.194 nm, respectively. We are not sure whether Si-C bonds formed or not; this is a matter open to interpretation.

\begin{figure}
\caption{Ball-and-stick models for CID 20149525. Left: initially generated by Open~Babel. Middle: initial chemical formula. Right: molecules calculated by PM6. The Si atom is considered to be in a mixture in the original, but is bonded in the PM6 optimized structure.} \label{CID:020149525}
\begin{tabular}{ccc}
 \includegraphics[width=5cm]{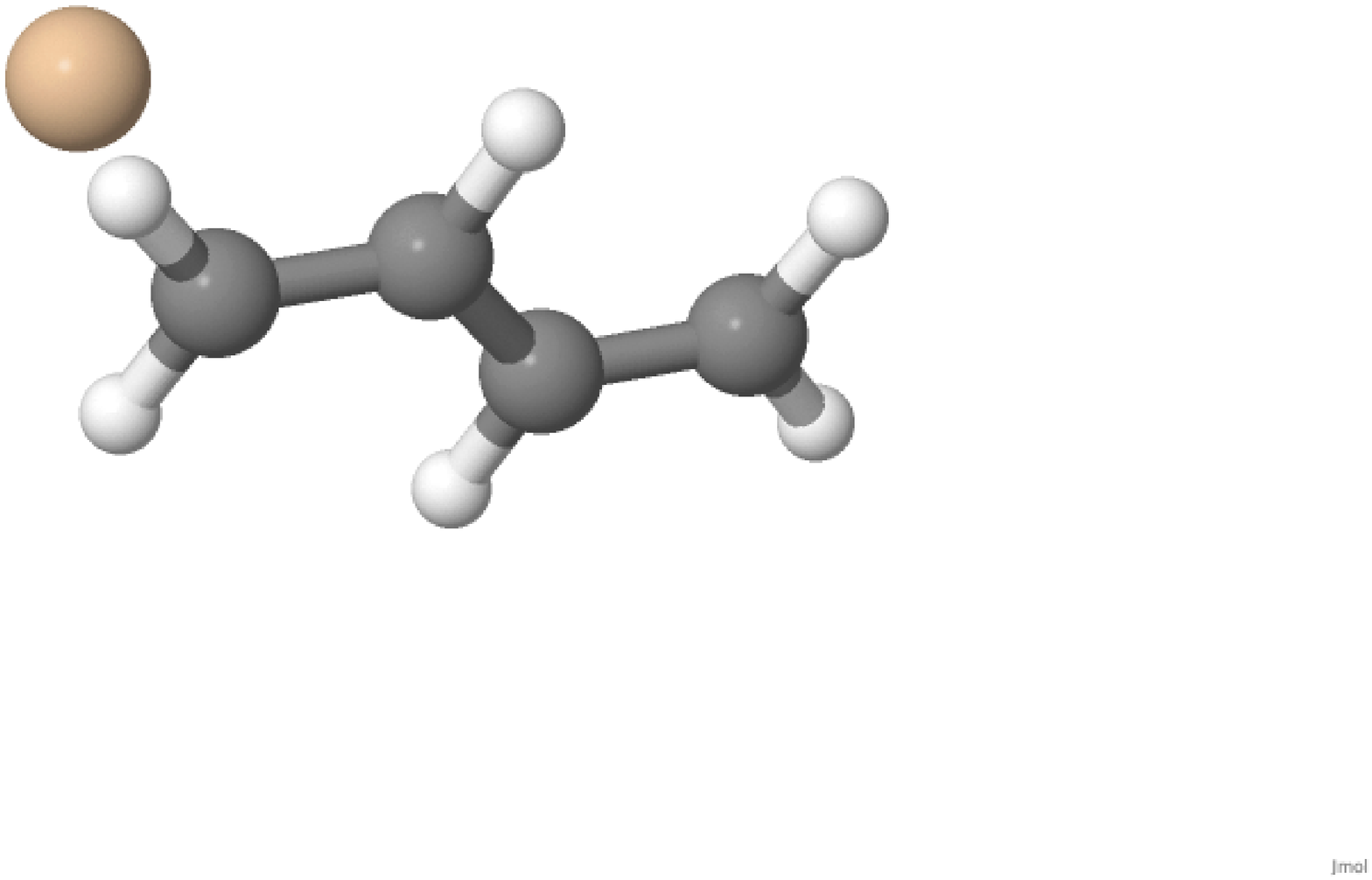} &
 \includegraphics[width=5cm]{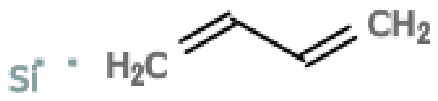} & 
 \includegraphics[width=5cm]{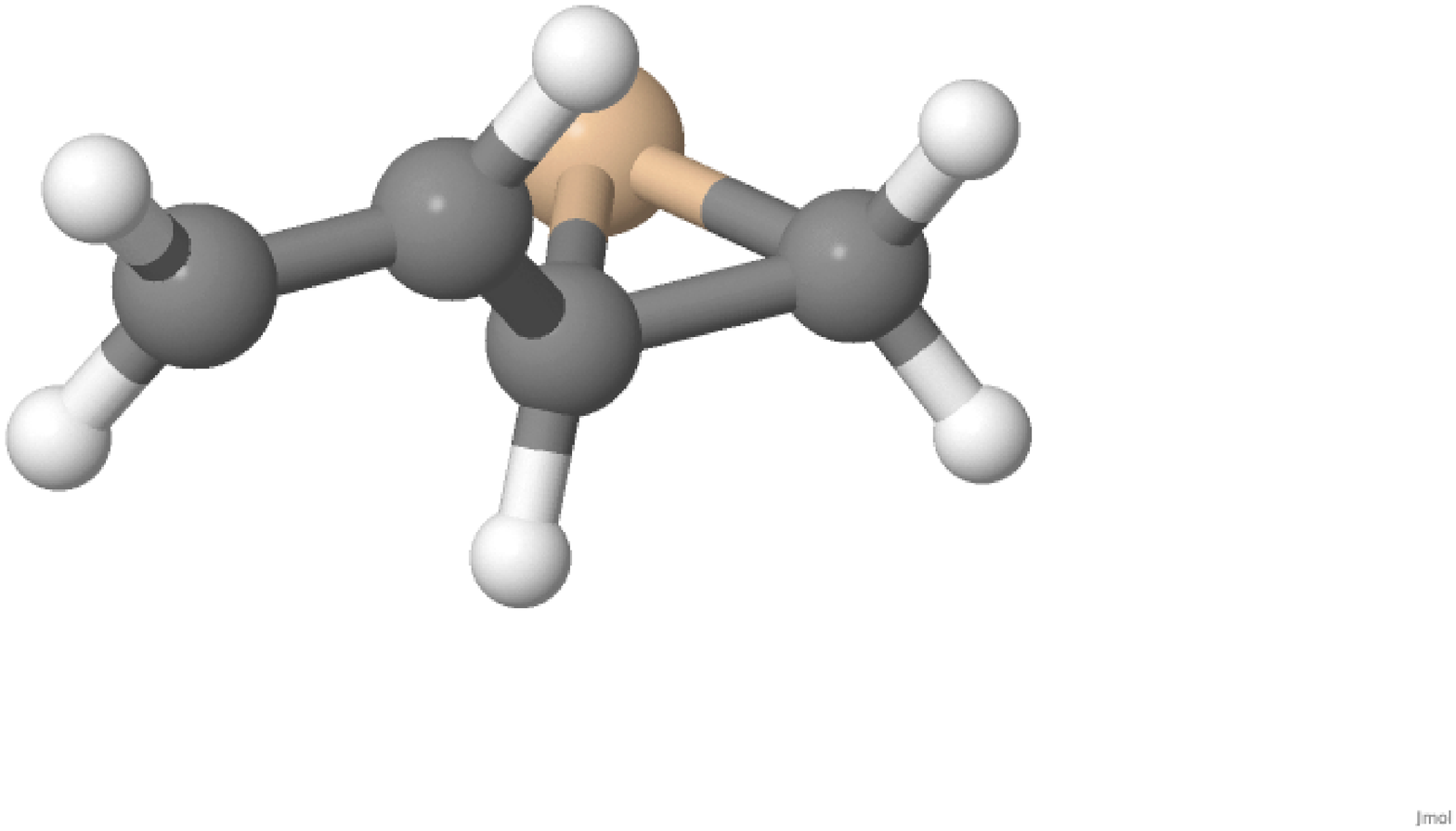} \\
\end{tabular}
\end{figure}

\item\textbf{Difference between molecular shapes for the same molecular InChI encoding. }\\
There are also awkward cases where the initial geometry guess looks wrong but the InChI is preserved (CID 53630746). Here, the initial InChI and InChI calculated from the initial geometry guess by Open~Babel are the same:
\begin{verbatim}
InChI=1S/C5H2O/c1-2-4-6-5-3-1/h4H2
\end{verbatim}

However, from the original 2D picture, the initial geometry generated by Open~Babel looks wrong. Fortunately, the InChI calculated by the PM6 optimized geometry is \texttt{InChI=1S/C5H2O/c1-4-2-6-3-5(1)4/h2H2}, which is different from the initial InChI. Detection of such cases is difficult.
\end{itemize}

\begin{figure}
\caption{Ball-and-stick models for CID 53630746. Left: initially generated by Open~Babel. Middle: initial chemical formula. Right: molecules calculated by PM6. The initial guess of Open~Babel failed to interpret the original intention. Problematically, the PM6 calculation did not fix this problem, or else the original molecule is unstable. } \label{CID:053630746}
\begin{tabular}{ccc}
 \includegraphics[width=5cm]{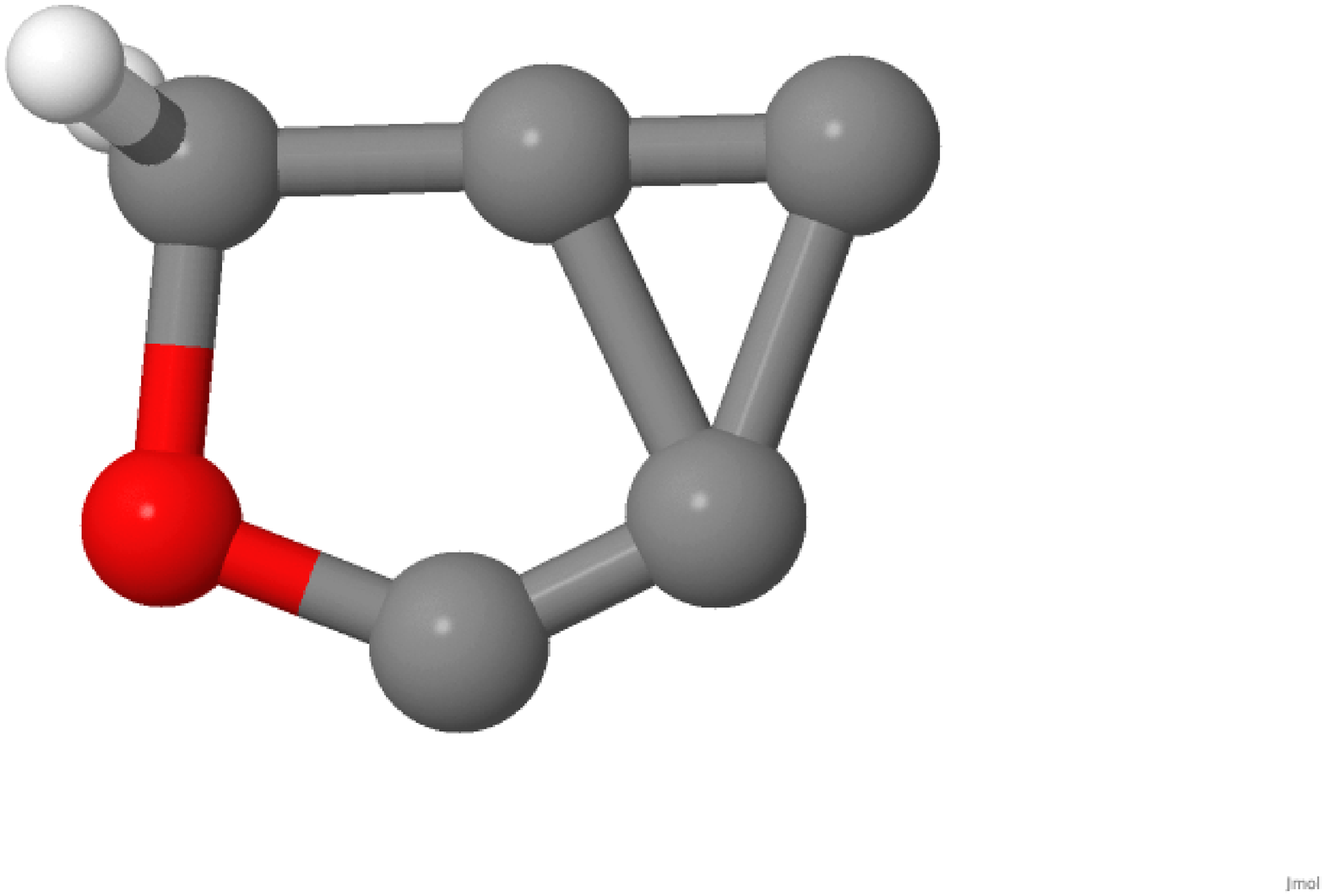} &
 \includegraphics[width=5cm]{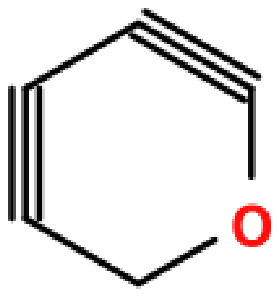} & 
 \includegraphics[width=5cm]{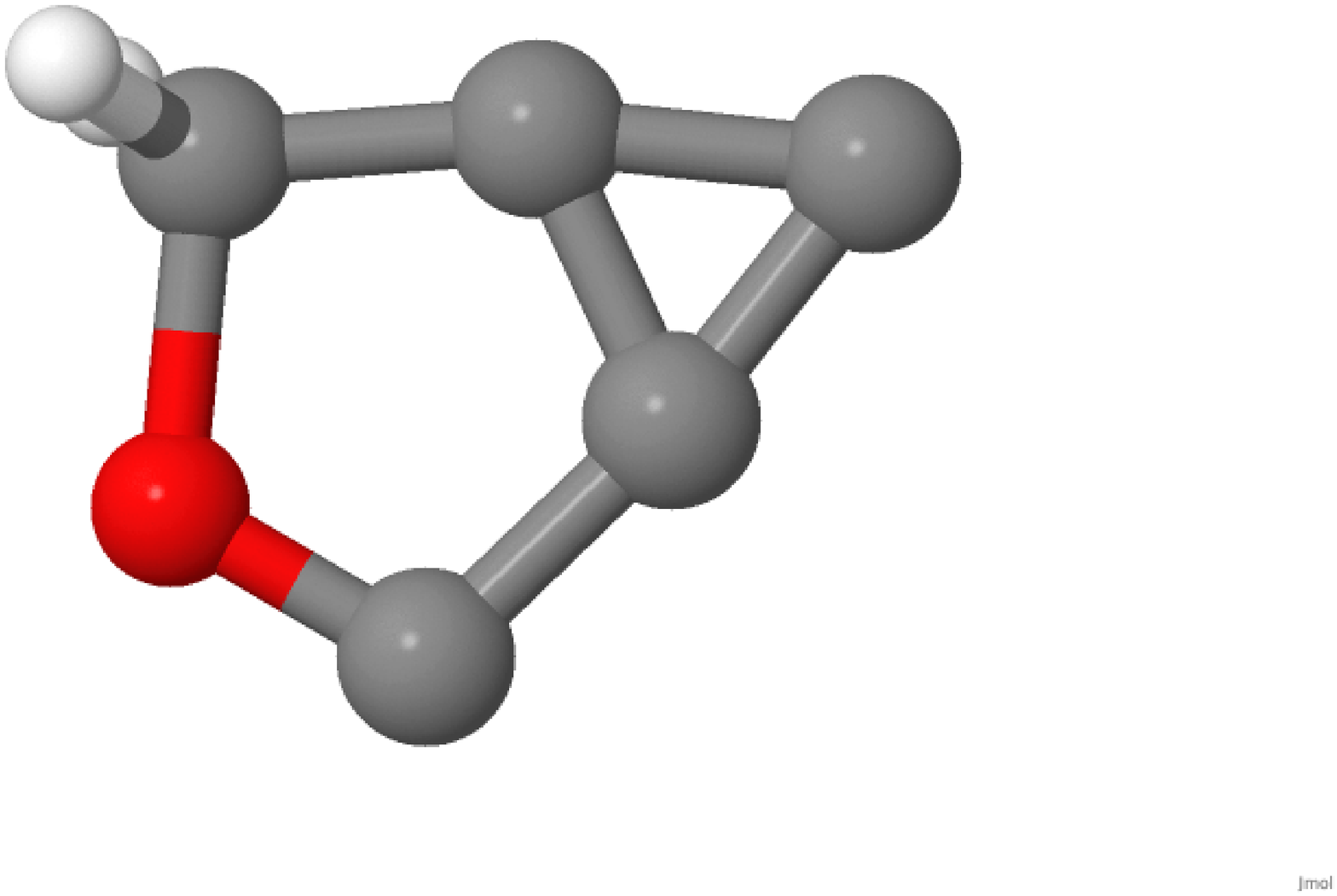} \\
\end{tabular}
\end{figure}

In our calculations, there were 1,816,733 molecules where the original and the PM6 optimized InChI were different in the connection layer. These issues are also related to  the detection or the generation of conformers of a molecule. The quantum chemical calculation can be used for curation in some cases.  Moreover, such an effort may improve molecular encoding methods. For example, we may add more layers to represent that CID 5987 and 60966 are stable when dissolved in water or when they are in the solid state.  For CID 53628168, we should add information indicating that this is an intermediate compound. 

\section{6. Future work}
We are preparing to publish more detailed analyses, in particular, on the optimization of the HOMO-LUMO gap, HOMO energy, LUMO energy, vibration intensity, modes, dipole moments, and structure changes, which should be useful for materials design. We are also planning to provide a database dump containing these data so that other researchers can easily make use of our results. Additionally, we have been running B3LYP calculations by using PM6 optimized geometries. A comprehensive investigation into failed calculations and curation of successfully calculated molecules will be necessary. Finally, we are also working on machine learning methods for quantitative structure–property relationships (QSPR) modelling  based on the dataset. The authors hope our contribution will help invigorate machine learning research on molecules.

\bibliography{Ronbun30}

\begin{acknowledgement}
The HOKUSAI facility was used to perform some of the calculations. This work was supported by the Japan Society for the Promotion of Science (JSPS KAKENHI Grant no. 18H03206).
\end{acknowledgement}

\end{document}